\documentclass[11pt]{article}
\pdfoutput=1
\usepackage[margin=1.0in]{geometry}
\usepackage{amsmath,amssymb,graphicx}
\usepackage[colorlinks=true, urlcolor=blue, linkcolor=blue]{hyperref}
\usepackage{braket}
\usepackage{csquotes}
\usepackage{float}

\usepackage{tikzducks}

\usepackage{framed}

\usepackage[font=small,labelfont=bf]{caption}
\usepackage{cite}

\newcommand{\be}{\begin{equation}}
\newcommand{\ee}{\end{equation}}

\usepackage{xcolor}
\usepackage{bm}

\title{\bf \large Lecture Notes for the 2023 Condensed Matter Summer School:\\ \ \\ \Large Introduction to random unitary circuits and the measurement-induced entanglement phase transition}
\author{Brian Skinner \\
{\small \color{gray} \texttt{skinner.352@osu.edu}}\\
{\small Ohio State University}}

\begin{document}

\maketitle

\begin{abstract}
There are several nice review articles about the physics of random unitary circuits and the measurement-induced entanglement phase transition, most notably these two: \cite{reviewrandomquantumcircuits2022,reviewentanglementdynamicsinhybridquantumcircuits}. This document has only two advantages over those nice reviews: (1) it's shorter, and (2) since this is not a real article, I am not bound by the usual conventions of scientific writing. \\ Look, a duck: \resizebox{5mm}{5mm}{ \tikz\duck;}.
\end{abstract}

\thispagestyle{empty}

\tableofcontents

\newpage

\setcounter{page}{1}

%\section*{\Large Part One: Entanglement entropy and random unitary circuits}
%\label{sec:lectureone}
%\addcontentsline{toc}{section}{\nameref{sec:lectureone}}

\section{Entanglement entropy and random unitary circuits}
\label{sec:differentialforms}

What distinguishes quantum information from classical information is the possibility of quantum entanglement. Generating, manipulating, and exploiting entanglement is sort of the whole point of quantum information as a science.

But entanglement can be a confusing quantity to think about: it doesn't correspond to any direct observable the way that, say, energy or electric charge do. So it can be tricky to answer basic statistical mechanical questions about quantum entanglement. Questions like:
\begin{itemize}
    \item How does entanglement grow or spread throughout a many body system with time?
    \item What are the statistics of random fluctuations in entanglement?
    %\item What happens to entanglement when someone is making measurements on the system?
\end{itemize}
These kinds of questions about dynamics are very natural in many settings in condensed matter physics. But applying them to quantum entanglement requires some effort to pose the questions properly. The random unitary circuit is one of the primary tools for posing and addressing these kinds of questions.

\subsection{How do you quantify quantum entanglement?}
\label{sec:Sdef}

The usual textbook definition of quantum entanglement is something like: 
\begin{displayquote}
``Quantum entanglement is when the state of the system cannot be completely specified by specifying the state of all its components in isolation."
\end{displayquote}
Or, if the textbook is being more mathematical, it will say something like:
 \begin{displayquote}
``Suppose that a quantum system contains two parts $A$ and $B$. If the state $\ket{\psi}$ of the system can be written as $\ket{\psi} = \ket{\psi_A} \otimes \ket{\psi_B}$, where $\ket{\psi_A}$ and $\ket{\psi_B}$ are states of $A$ and $B$ separately, then there is no quantum entanglement between $A$ and $B$. If $\ket{\psi}$ cannot be written this way, then $A$ and $B$ are entangled.''
\end{displayquote}

So, for example, if I have two spins $A$ and $B$, then the state 
\be 
\ket{\psi}_\text{unentangled} = \left( \frac{\ket{\uparrow_A} + \ket{\downarrow_A} }{ \sqrt{2} } \right) \otimes \left( \frac{\ket{\uparrow_B} - \ket{\downarrow_B} }{ \sqrt{2} } \right) = \frac12 \left( \ket{\uparrow_A \uparrow_B} - \ket{\uparrow_A \downarrow_B} + \ket{\downarrow_A \uparrow_B} - \ket{\downarrow_A \downarrow_B} \right)
\label{eq:psiunentangled}
\ee 
is one with no entanglement between the two spins, since the state can be fully specified by saying that spin $A$ is in the state $(\ket{\uparrow} + \ket{\downarrow})/\sqrt{2}$ and spin $B$ is in the state $(\ket{\uparrow} - \ket{\downarrow})/\sqrt{2}$. 

On the other hand, the singlet state
\be 
\ket{\psi}_\text{singlet} = \frac{1}{\sqrt{2}} \left( \ket{\uparrow_A \downarrow_B} - \ket{\downarrow_A \uparrow_B} \right)
\ee 
\emph{is} entangled, since this state cannot be described as (something for A) $\times$ (something for B).

The annoying thing about this usual definition, though, is that it gives you the impression that entanglement is a kind of binary thing: that you either can or cannot write the wave function as a product. In other words, it gives you no ability to think quantitatively about the question \emph{how entangled} are $A$ and $B$? So I personally prefer to think about entanglement in terms of measurement outcomes, and to what degree making measurements on part B will affect the subsequent likelihood of getting different measurement outcomes for A.

Notice, for example, that if I have the state $\ket{\psi}_\text{unentangled}$ written above and I measure the value of spin B, then I will have two equally-likely outcomes: $\uparrow$ or $\downarrow$. If I subsequently measure the state of spin A, after collapsing the state of spin B, then I will also have two equally likely outcomes. But the same would have been true if I had never done any measurements of spin B. So measuring the state of spin B does not at all affect how likely I am to get different measurement outcomes for A. Thus I can conclude that A and B are unentangled.

Contrast this with what happens for the singlet state $\ket{\psi}_\textrm{singlet}$. If I first measure B, then I can get one of two equally likely outcomes: $\uparrow$ or $\downarrow$. Let's suppose I measure B to be $\downarrow$. Once I make this measurement for B, then I \emph{collapse} the wave function to be $\ket{\psi} = \ket{\uparrow_A \downarrow_B}$. 
A subsequent measurement of spin A will definitely yield $\uparrow$; there is no longer any uncertainty about what the measurement will give. In other words, by first measuring B, I have reduced (in this case, entirely) the \emph{statistical entropy} of the measurement outcomes of A.

This is the basic idea behind the quantity called the \emph{entanglement entropy} $S$. It quantifies the degree to which the statistical entropy of measurement outcomes for A is reduced by first completely measuring the state of B. To define $S$, we first construct the density matrix $\rho$. For a pure state $\ket{\psi}$,
\be 
\rho = \ket{\psi} \bra{\psi}.
\ee 
(In these lecture notes I will mostly be describing pure states, which are described by a wave function, as opposed to ``mixed states", which are a statistical sum of different wave functions.)
I can define the ``reduced density matrix'' for A, $\rho_A$, by doing a partial trace over values of B:
\be 
\rho_A = \textrm{Tr}_B (\rho).
\ee 
The ``partial trace'' means partitioning the density matrix $\rho$ into sectors according to the value of $A$, and the doing the matrix trace within each sector. So, for example, the density matrix for the singlet state is
\be 
\rho_\text{singlet} = \frac12 
\begin{pmatrix}
0 & 0 & 0 & 0 \\
0 & 1 & -1 & 0 \\
0 & -1 & 1 & 0 \\
0 & 0 & 0 & 0
\end{pmatrix},
\ee
where the four rows/columns correspond to $\uparrow_A \uparrow_B$, $\uparrow_A \downarrow_B$, $\downarrow_A \uparrow_B$, and $\downarrow_A \downarrow_B$. 
The partial trace corresponds to doing a trace (sum of the diagonal) across the $2 \times 2$ sub-matrices in each corner. This procedure gives
\be \rho_A = \textrm{Tr}_B (\rho) = \frac12
\begin{pmatrix}
1 & 0\\
0 & 1
\end{pmatrix}.
\ee 
Notice that this $\rho_A$ is not something that could have been written as the outer product of any wave function with itself -- it cannot represent any kind of pure state. Instead, it looks like a statistical mixture: 50\% $\uparrow$ and 50\% $\downarrow$. The statistical entropy in this reduced density matrix tells you how entangled A is with B. (A quantum pure state always has zero statistical entropy -- it is in a specific quantum state, rather than a probabilistic mixture of different states.)

So now we define the \emph{entanglement entropy} as
\be 
S_1 = - \textrm{Tr}(\rho_A \log \rho_A).
\label{eq:SvN}
\ee 
The right-hand side should remind you of the Shannon entropy of a probability distribution
\be 
S_\textrm{Shannon} = - \sum_i p_i \log(p_i),
\ee 
where $i$ labels different states. Taking the log of a matrix can be confusing, so it is often best to think about Eq.~\ref{eq:SvN} in terms of the eigenvalues $\lambda_i$ of the reduced density matrix $\rho_A$:
\be 
S_1 = - \sum_i \lambda_i \log(\lambda_i).
\ee 
(It is a general properties of density matrices that they are diagonalizable and have eigenvalues that sum to 1, $\sum_i \lambda_i = 1$, since the diagonal elements of the density matrix correspond to the probability of getting a particular measurement outcome.)

As it turns out, there is an entire family of similar measures of entropy called the ``Renyi entropies'' $S_n$. These are all similar, but involve raising the probabilities to different powers $n$. Specifically,
\be 
S_n = \frac{1}{1 - n} \log \left( \textrm{Tr}( \rho_A^n)  \right) = \frac{1}{1 - n} \log \left( \sum_i \lambda_i^n \right).
\label{eq:Sn}
\ee 
These Renyi entropies are not usually very different from each other conceptually, so it's not typically worth making a big deal about the distinction between different values of $n$. But it's worth singling out three cases: $n \rightarrow 1$, $n \rightarrow 0$, and $n \rightarrow \infty$.

The limit $n \rightarrow 1$ of Eq.~\ref{eq:Sn} has to be taken carefully, but if you do you'll get Eq.~\ref{eq:SvN}, which is usually called the ``von Neumann entanglement entropy". It is perhaps the most common measure of entanglement entropy.

In the limit $n \rightarrow 0$, any nonzero eigenvalue of $\rho_A$ becomes raised to the $0$th power and thus becomes equal to $1$, so we can write
\be 
S_0 =  \log( \textrm{\# of nonzero eigenvalues of } \rho_A ).
\ee 
$S_0$ gets called the ``Hartley entropy'' or ``zeroth Renyi entropy".

$S_0$ is a kind of stupid quantity in the sense that it changes discontinuously after an infinitessimal interaction. Imagine, for example, two unentangled spins. If I bring them together and allow them to interact for a very short time, I might expect that any entanglement between them will be very small. But $S_0$  jumps immediately from $0$ to $\log 2$, and in that sense $S_0$ is kind of a binary quantity that distinguishes only between ``some entanglement'' and ``no entanglement'' (I make this statement a bit more precise in the next subsection). All other $S_n$ with $n > 0$ increase infinitesimally as you would expect. But, as we'll see below, the Hartley entropy $S_0$ is useful in that it can usually be thought about in a straightforward classical way. Notice also that 
\be 
S_0 \geq S_n \textrm{  for all } n \geq 0 .
\ee 
That is, $S_0$ provides an upper bound for all the other $S_n$'s.

In the limit $n \rightarrow \infty$, the sum in Eq.~\ref{eq:Sn} is dominated by the single largest eigenvalue $\lambda_i$, and so
\be 
S_\infty = \log(1/\lambda_\textrm{max}).
\ee 
The nice thing about $S_\infty$ is that it provides both an upper and lower bound for all other $S_n$'s with $n > 1$. It turns out that \cite{wilming_entanglement_2019}
\be 
S_\infty \leq S_n \leq \frac{n}{n-1} S_\infty \textrm{  for } n > 1.
\label{eq:Sineq}
\ee 
So, for example, if some $S_n$ with $n > 1$ diverges, then all other $S_n$ with $n > 1$ also diverge; and if some $S_n$ with $n > 1$ remains small then so do all the others. So we often talk about a ``generic'' behavior of $S_n$ corresponding to $n > 1$. In principle Eq.~\ref{eq:Sineq} says nothing about $n = 1$, but in all cases that I know of $S_1$ behaves similarly to the ``generic'' case. By contrast, there are no guarantees that $S_0$ will behave in the ``generic'' way.

\subsection{The Schmidt decomposition}
\label{sec:Schmidt}

There is a useful and general way to represent an entangled state called the \emph{Schmidt decomposition}. The idea is that for any quantum state $\ket{\psi}$ on subsystems A and B there is some set of orthonormal states $\{\ket{i_A}\}$ and $\{\ket{i_B} \}$ such that $\ket{\psi}$ can be written
\be 
\ket{\psi} = \sum_i \gamma_i \ket{i_A} \otimes \ket{i_B}
\label{eq:decomposition}
\ee 
for some coefficients $\gamma_i$. 

Notice that there are in general multiple ways to write a state $\ket{\psi}$ as a sum over terms like $\ket{i_A} \otimes \ket{i_B}$. For example, the state $\ket{\psi}_\textrm{unentangled}$ that we looked at on page 1 could be written in a smart way as a sum with only one term (choosing the states $\ket{i_A}$ and $\ket{i_B}$ as in the first equality of Eq.~\ref{eq:psiunentangled}); or it could be written in a dumb way as a sum over four terms: 
$\ket{\uparrow_A} \otimes \ket{\uparrow_B}$, $\ket{\uparrow_A} \otimes \ket{\downarrow_B}$, $\ket{\downarrow_A} \otimes \ket{\uparrow_B}$, and $\ket{\downarrow_A} \otimes \ket{\downarrow_B}$.

The Schmidt decomposition generally refers to the smartest way to do things, with the minimal number of terms in the sum. In fact, in this optimally compact decomposition the coefficients $\gamma_i$ are the square root of the eigenvalues $\lambda_i = \gamma_i^2$ of the reduced density matrix $\rho_A$. 

One implication is that the zeroth Renyi entropy $S_0$ is exactly given by the (log of the) number of terms in the Schmidt decomposition:
\be 
S_0 =  \log( \textrm{\# of nonzero terms in the Schmidt decomposition of } \ket{\psi} ).
\label{eq:S0Schmidt}
\ee 
That is, $S_0$ counts the minimal number of terms that you need in order to write a decomposition of the form of Eq.~\ref{eq:decomposition} if you want to represent the state exactly. But remember that the Schmidt decomposition is the smart way to write a state, and there are generally many dumb decompositions with more terms. So a corollary of Eq.~\ref{eq:S0Schmidt} is
\be 
S_n \leq S_0 \leq \log( \textrm{\# of nonzero terms in \emph{any} decomposition of } \ket{\psi} ).
\label{eq:S0any}
\ee 
This last equation will be important when we discuss the idea of the minimal cut in Sec.~\ref{sec:mincutsec}.

\subsection{The unitary circuit and the random unitary operator}
\label{subsec:exteriorderiv}

\subsubsection{The unitary circuit}

\begin{figure}[!h]
\centering
\includegraphics [width = 0.9\textwidth]{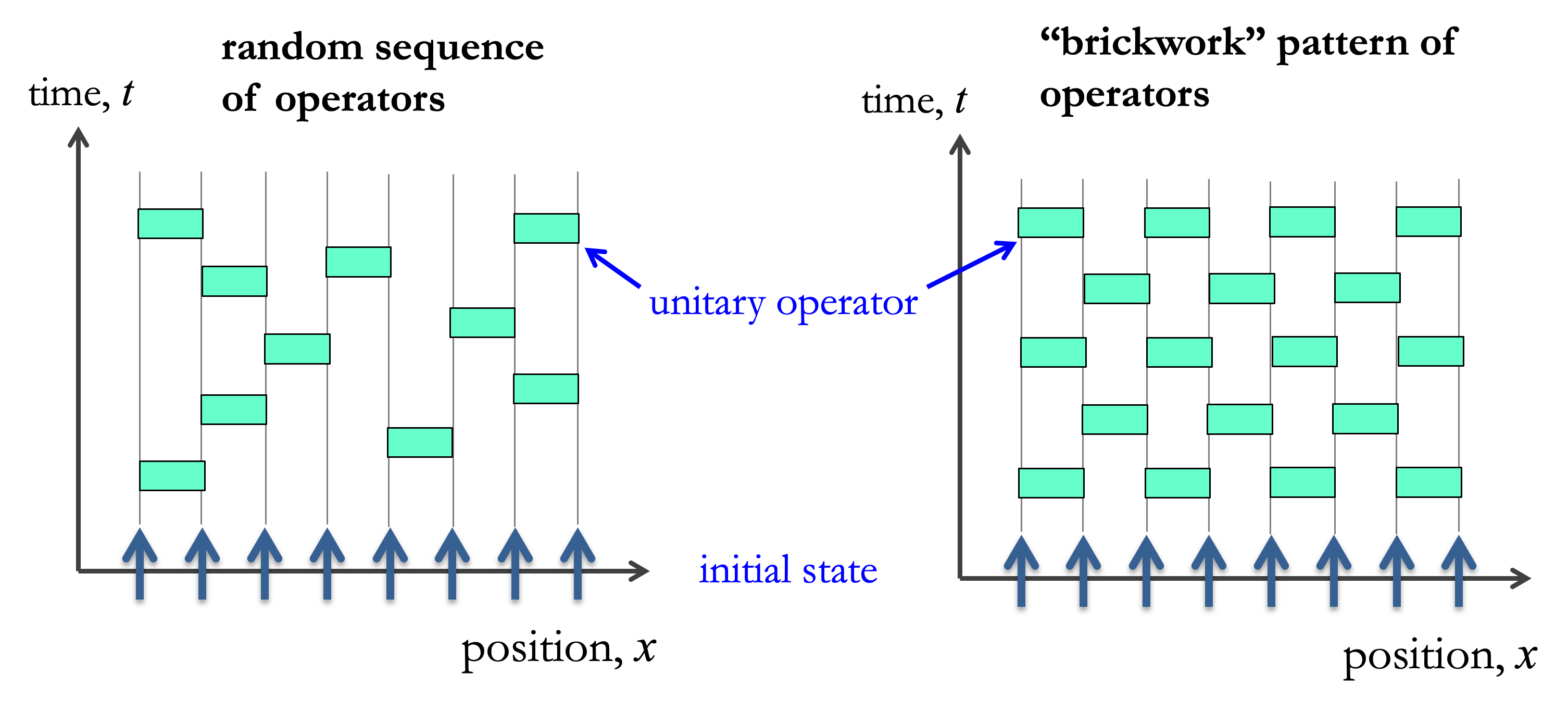}
\caption{Examples of unitary quantum circuits, which evolve an initial state (here represented by a chain of $\uparrow$ spins) via a sequence of discrete unitary operators. In this example the operators act on two neighboring spins, and are arranged in either a random (left) or uniform (right) pattern. My apologies for drawing time in the vertical direction; I know that real quantum information people tend to draw their circuits with time going to the right.
} \label{fig:circuits}
\end{figure}

A ``unitary circuit'' is a way of describing the evolution of a quantum state in discrete time rather than continuous time. In a unitary circuit, the quantum state evolves through the application of discrete operators to the quantum state (see Fig.~\ref{fig:circuits}).

There are multiple reasons why you might want to describe time evolution with a quantum circuit, rather than through a continuous-time evolution operator like $U = \exp(-i H t / \hbar)$. For example, maybe you are trying to describe a quantum computer, which operates using discrete quantum gates. Or maybe you really do care about a system with a continuous-time Hamiltonian, but for reasons of numerical or conceptual convenience you would like to approximate the dynamics by a sequence of discrete operators. For example, the picture on the right of Fig.~\ref{fig:circuits} could serve as an approximation of a 1D spin chain with nearest-neighbor interactions.

Actually, there is a general theorem (often called the Lie-Trotter formula) that guarantees the validity of approximating any continuous time Hamiltonian via a sequence of discrete operators. Suppose I have a Hamiltonian $H$ that is a sum of many short-ranged parts $H_j$, such that ${H = \sum_j H_j}$. In general the time evolution operator cannot be written as a simple product of separate operators for each $H_j$. I mean that
\be 
e^{-i H t / \hbar} \neq e^{-i H_1 t / h} \times e^{-i H_2 t / \hbar} \times e^{-i H_1 t / h} \times ...
\ee 
Sadly, that's not how matrix multiplication works. But the above expression becomes \emph{almost} correct if the time interval $\Delta t$ of the evolution is very short. That is:
\be 
e^{-i H (\Delta t)/\hbar} = e^{-i H_1 (\Delta t)/\hbar} \times e^{-i H_2 (\Delta t)/\hbar} \times e^{-i H_3 (\Delta t)/\hbar} \times ...  + O(\Delta t)^2.
\ee 
In other words, the time evolution of the full, many-body system can always be approximated by a product of many local operators $\exp(-i H_j (\Delta t)/\hbar)$, so long as I divide up the total time evolution into many small increments $\Delta t$. Since the error from this approximation in a single step grows like $(\Delta t)^2$, the total error from a long time evolution $t$ is still of order $O(\Delta t/t)$. So the dynamics of any many-body quantum system can be ``Trotterized'' into a quantum circuit, so long as you have the patience to put enough layers of operators in your circuit. And some circuit like the one on the right side of Fig.~\ref{fig:circuits} starts to look like a universal approximation of a system with nearest-neighbor interactions.

\subsubsection{Aside: who cares about random dynamics?}

For people interested in the statistical mechanics of quantum entanglement growth, a big point of emphasis is the \emph{random} unitary circuit. A random circuit is one for which either the structure of operators in time in space, or the operators themselves, are chosen in some random way. 
The very notion of ``random dynamics'' can seem a bit perverse -- after all, no one operating a quantum computer or trying to figure out the ground state of a many-body Hamiltonian is thinking about random dynamics. And when the dynamics is random in time, it means that there is no concept of a Hamiltonian and no concept of energy: one cannot talk about ``ground states'' or ``excited states." 

But there are reasons why one might want to think about random dynamics nonetheless. First, there is the hope that entanglement dynamics under a random circuit will be \emph{generic} in some way. Just as the properties of random matrices end up being useful descriptors of a whole category of Hamiltonians, we can hope that by abstracting away from any specific Hamiltonian to a random quantum circuit we can uncover features of entanglement dynamics that are generic to a broad class of situations. Then, of course, specific situations might be ``non-generic'' in a variety of different ways -- most likely, by introducing some conserved quantity (the total value of $S_z$, for example). For the random dynamics considered in these lecture notes, it's important that there are no conserved quantities.

Part of the motivation for studying entanglement dynamics in this way comes from the physics of thermalization. The question of thermalization concerns whether a quantum system that is completely isolated from its environment will be able to ``act as its own bath'', such that quantities like charge, energy, etc.\ will be able to diffuse around and fill space in the usual way. One of our primary perspectives on thermalization is the Eigenstate Thermalization Hypothesis (ETH), which can be stated as follows. Consider a very large system, which we will divide into a macroscopic subsystem A and its (macroscopic) complement B. Suppose that the entire system is in an energy eigenstate $\ket{\psi}$ with energy $E$. One can associate a temperature $T$ with this eigenstate by saying that one would get the same energy (in expectation value) from a classical Boltzmann distribution if the temperature had a specific value $T$. Such a classical Boltzmann distribution has a corresponding density matrix $\rho_{eq}(T)$ that is boring and classical: its diagonal entries are proportional to $\exp(-E_i/k_BT)$, where $E_i$ is the energy of the eigenstate $i$, and the off-diagonal entries of $\rho_{eq}(T)$ are all zero. The ETH asserts that, if the energy eigenstate is ``thermalized'', then the \emph{reduced density matrix} $\rho_A$ for the subsystem A satisfies\footnote{The last equality in Eq.~\ref{eq:ETH} makes use of the large system size limit -- in general the off-diagonal components of the reduced density matrix $\rho_A$ are not literally zero under ETH, and there are interesting things to be said about them. But they are much smaller in magnitude than the diagonal components when A and its complement are both large.}
\be 
\rho_A = \textrm{Tr}_B( \rho ) = \rho_{eq}^A(T).
\label{eq:ETH}
\ee 
In other words, tracing out everything outside of the region A leaves you with something that looks like the usual classical thermodynamics. (See \cite{Nandkishore2015, deutsch_eigenstate_2018} for reviews of ETH.)

But this statement of ETH can only be satisfied if the subsystem A has extensive entanglement with its environment, since an equilibrium distribution at finite temperature has extensive entropy. We know of many systems that satisfy ETH (and in some sense satisfying ETH is the norm; we tend to find violations of ETH more interesting). So the question of how a system \emph{thermalizes} -- how it goes from something with little entanglement to something with a sufficiently extensive entanglement to satisfy ETH -- is a question about entanglement dynamics. The random unitary circuit provides us with a tool to study some kind of ``generic entanglement dynamics'', where entanglement forms more or less as quickly as allowed by spatial locality. Of course, in some particular situation this generic dynamics may or may not be stymied by the formation of some conserved quantity that bottlenecks the growth of entanglement.

Studying how quickly entanglement grows (or doesn't) can be important for another more pragmatic reason. Until quantum computers become as good as Michio Kaku imagines them to be, simulating quantum systems is difficult computationally. And the more entangled the system is, the harder it is to simulate. Quantum states with only short-range entanglement have an efficient classical representation (say, in terms of matrix product states), but states with extensive entanglement do not. So knowing whether or not your state is going to develop extensive quantum entanglement becomes crucial for knowing whether your code is going to finish running in time for you to graduate.

\subsubsection{Random unitary operators}

The most generic type of random unitary operator is what's called a ``Haar-random'' matrix. A Haar-random matrix is one that is chosen uniformly at random from the space of all possible unitary matrices.

Haar-random matrices are in some sense the ``most random'' operators possible, so they have been a major point of study. But they have the drawback that they necessarily produce dynamics that is hard to simulate classically: over time any nice and easily-written product state will spread to fill the full $2^N$-sized Hilbert space (say, of $N$ qubits\footnote{Throughout these notes I'm going to use the words ``spin'' and ``qubit'' interchangeably. Sorry.}), so that even storing the wave function becomes difficult computationally.

For that reason a common alternative is to instead consider only a more limited set of operators called Clifford operators. Clifford operators have the advantage that they shift the quantum state around within a limited subset of the Hilbert space, such that the state always has a compact classical representation (called a ``codeword''). (See Sec.~3.3.4 of Ref.~\cite{reviewrandomquantumcircuits2022}  for a nice and short overview of the Clifford group.) In this sense Clifford dynamics is ``less random'', but it allows one to simulate large system sizes ($\sim 10^3$ qubits) and long times on ordinary computers, while Haar-random dynamics is usually limited to only a couple dozen qubits.

\subsection{The minimal cut: a classical bound on entanglement}
\label{sec:mincutsec}

Quantum mechanics is hard [{\color{green}18}-{\color{green}19,654}]. For that reason, much of the important progress in the field of entanglement dynamics has proceeded by finding clever analogies or mappings to classical problems. One of the cleverest and most central of these ideas (mostly developed in Ref.~\cite{Nahum_Quantum_2017}) is what's  called the ``minimal cut''. The minimal cut is the most prominent recurring theme of these lecture notes.

\subsubsection{Heuristic argument for the minimal cut}
\label{sec:mincut}

To understand, at a gut level, the minimal cut idea, imagine a circuit that looks like this:
\begin{figure}[H]
\centering
\includegraphics [width = 0.35\textwidth]{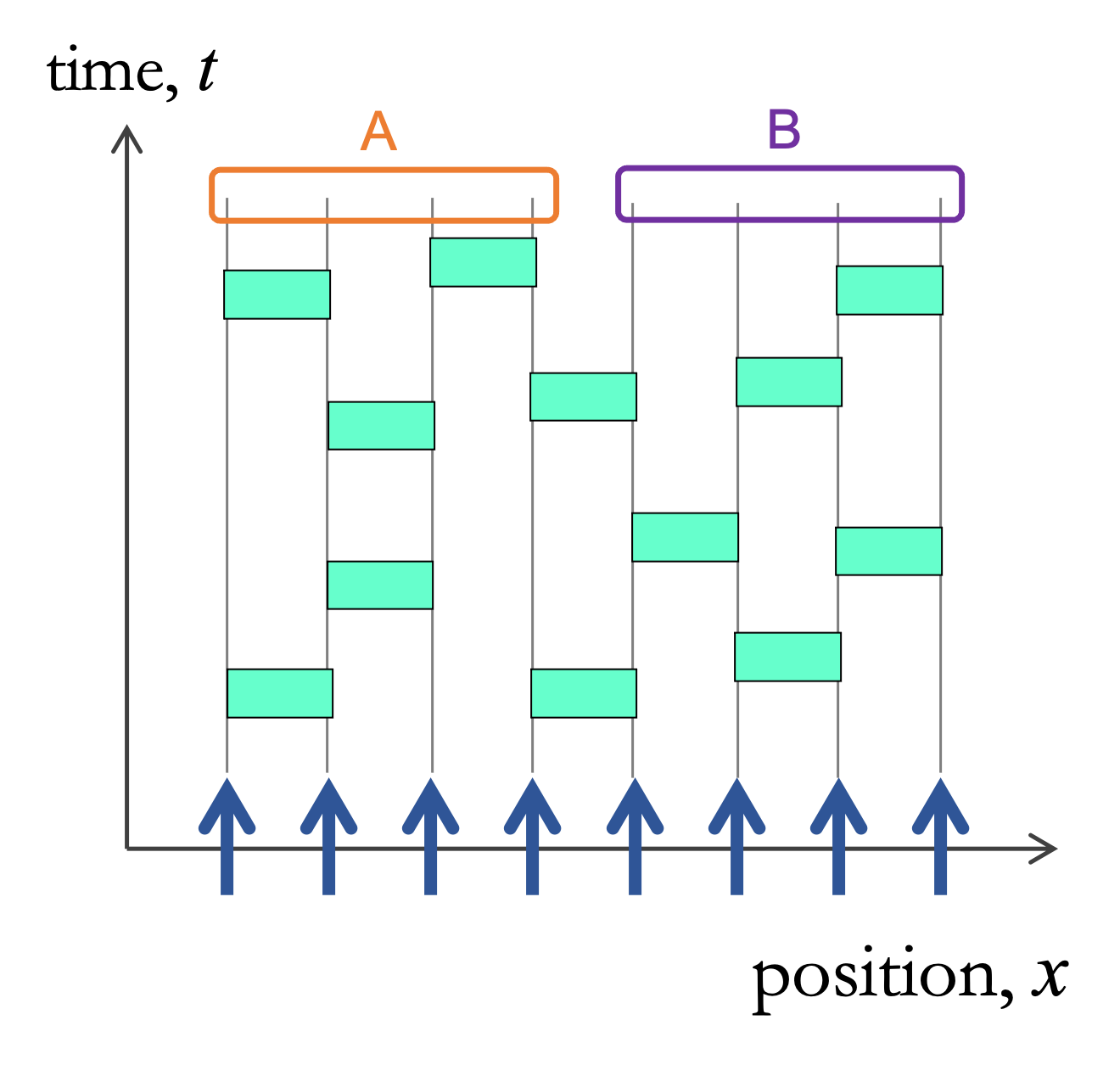}
\caption{A quantum circuit that transforms an initial state into a final state with two parts, A and B.
} \label{fig:circuitnocut}
\end{figure}
This circuit can be viewed as a kind of machine (formally, a tensor network) that takes a given input state to a corresponding output state. 

Now suppose I want to estimate the entanglement entropy between the left half of my system and the right half (labeled A and B above) at the final time. I can, arbitrarily, draw a red line that divides this one tensor network into two smaller tensor networks, one whose output contains the state of the A qubits at the final time, and another whose output contains the state of the B qubits at the final time. Like so, perhaps:

\begin{figure}[H]
\centering
\includegraphics [width = 0.9\textwidth]{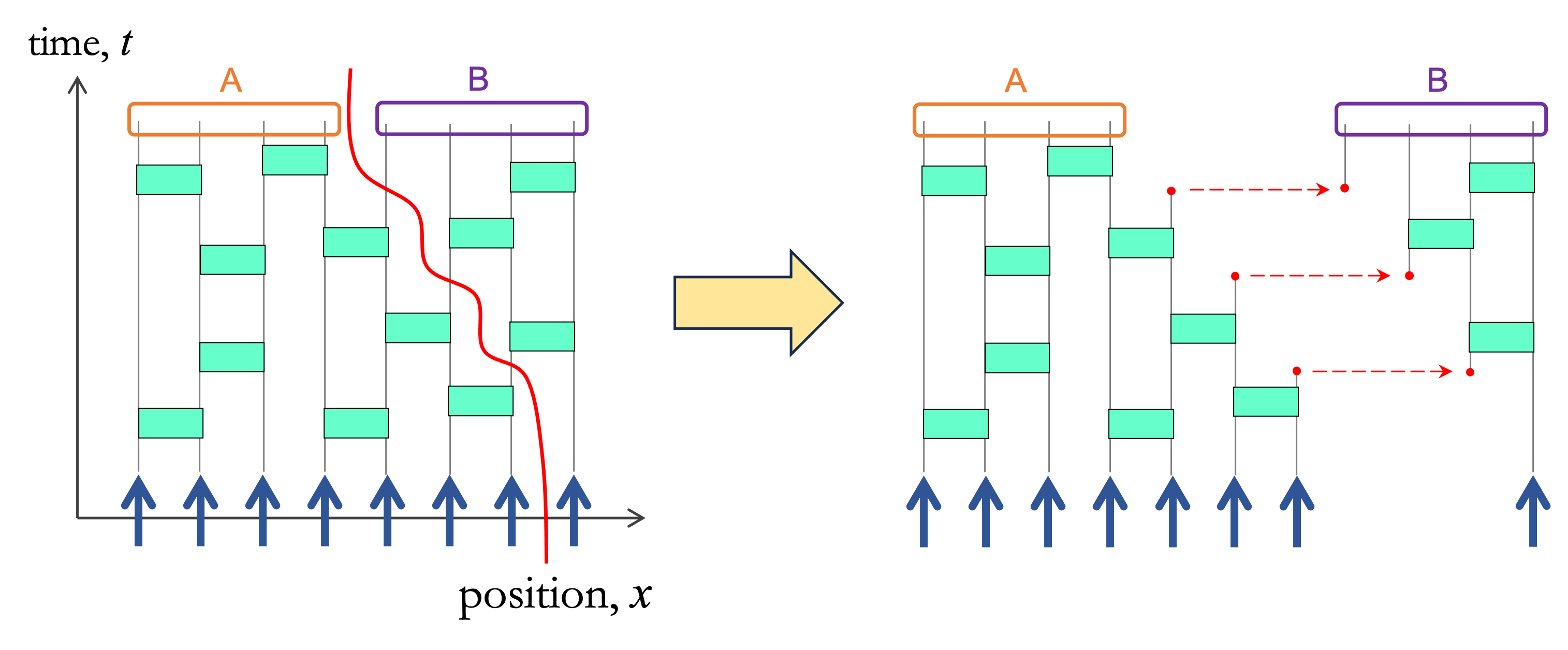}
\caption{I can partition my circuit into two tensor networks that output the states of A and B respectively. In order to construct the full circuit I need to feed some of the outputs from the left tensor network (denoted by red dots) into the right tensor network.
} \label{fig:circuitwcut}
\end{figure}

But by making this partition I have also created additional inputs and outputs for the two tensor networks. Here, the states of the qubits highlighted in red are outputs of the left tensor network that are used as inputs for the right tensor network. Each of these red circles can transmit at most one bit of information between the two tensor networks. In this example there are three shared connections between the two tensor networks, so consequently there can be at most three bits of information shared, which means that the entanglement entropy between A and B at the final time must be less than $3 \times \log 2$.

More generally, define $N_\textrm{cut}$ as the number of legs of the circuit (black lines) that are severed when I draw a red line partitioning the tensor network into two pieces. (My line must always start at the top of the circuit at the midpoint between subsystems A and B, so that the output states of the two smaller tensor networks fully contain the subsystems A and B.) The quantity $N_\textrm{cut}$ is equal to the number of shared connections between the two tensor networks. Since each connection can transmit at most one bit of information, I am generally guaranteed that for any $S_n$
\be 
S_n \leq N_\textrm{cut} \times \log 2.
\label{eq:Sncut}
\ee 

A more precise way to arrive at Eq.~\ref{eq:Sncut} is to realize that the partition shown in Fig.~\ref{fig:circuitwcut} actually represents a proposal for a way to write the final state $\ket{\psi}$ as a decomposition. Let's use $U_A$ to denote the operator defined by the left tensor network; in this example $U_A$ takes an input state with 7 $\uparrow$ spins to a state of 7 other spins (four that belong to A and three others that will be fed as inputs to the right tensor network).  Meanwhile, the right tensor network is an operator $U_B$ that takes the state of four spins (three outputs from the left plus one $\uparrow$ from the initial time) to the four spins in B. Now notice that I can write the final state as
\be 
\ket{\psi} = \sum_{\sigma_A, \sigma_B} \sum_{\sigma_\textrm{shared}} a(\sigma_A, \sigma_\textrm{shared}) \ket{\sigma_A} \otimes b(\sigma_B, \sigma_\textrm{shared}) \ket{\sigma_B}.
\label{eq:cutdecomposition}
\ee 
Here, the sum $\sum_{\sigma_A, \sigma_B}$ represents a sum over all possible states of the spins at the final time. And the sum $\sum_{\sigma_\textrm{shared}}$ is a sum over all possible states of the spins that are shared between the two tensor networks. The coefficients $a(\sigma_A, \sigma_\textrm{shared})$ and $b(\sigma_B, \sigma_\textrm{shared})$ represent the matrix elements of the operators $U_A$ and $U_B$:
\begin{align}
a(\sigma_A, \sigma_\textrm{shared}) & = \bra{\sigma_A, \sigma_\textrm{shared}} U_A \ket{\uparrow \uparrow \uparrow \uparrow \uparrow \uparrow \uparrow}, \\
b(\sigma_B, \sigma_\textrm{shared}) & = \bra{\sigma_B} U_B \ket{\sigma_\textrm{shared}, \uparrow}.
\end{align}

There is nothing clever about Eq.~\ref{eq:cutdecomposition}; I am just summing directly over all possible output basis states of $U_A$ and $U_B$, keeping in mind that the shared spins $\sigma_\textrm{shared}$ are the same. But notice that I can rearrange Eq.~\ref{eq:cutdecomposition}  slightly to get this:
\be 
\ket{\psi} =  \sum_{\sigma_\textrm{shared}} \left( \sum_{\sigma_A} a(\sigma_A, \sigma_\textrm{shared}) \ket{\sigma_A} \right) \otimes \left(\sum_{\sigma_B} b(\sigma_B, \sigma_\textrm{shared}) \ket{\sigma_B} \right).
\label{eq:cutdecompositionrearranged}
\ee 
Now you can see that I am proposing a way to write the state $\ket{\psi}$ as a decomposition in the style of Eq.~\ref{eq:decomposition}. The two quantities in the big parentheses represent state $\ket{i_A}$ and $\ket{i_B}$ that live only on the subsystems A and B. The number of terms in this proposed decomposition is the same as the number of possible states of the shared spins; so for this example my proposed decomposition has $2^3 = 8$ terms. Now you can recall Eq.~\ref{eq:S0any}, which tells me that my partition gives an upper bound on all the entropies:
\be 
S_n \leq \log \left( 2^{N_\textrm{cut}} \right) = N_\textrm{cut} \log 2.
\ee

Now comes the clever part. If you stare at Fig.~\ref{fig:circuitwcut} for a moment you'll realize that there is a more frugal way to divide this circuit into two pieces. I mean that I could have drawn a partition that crosses fewer legs of the circuit. This partition involves only one shared connection:

\begin{figure}[H]
\centering
\includegraphics [width = 0.35\textwidth]{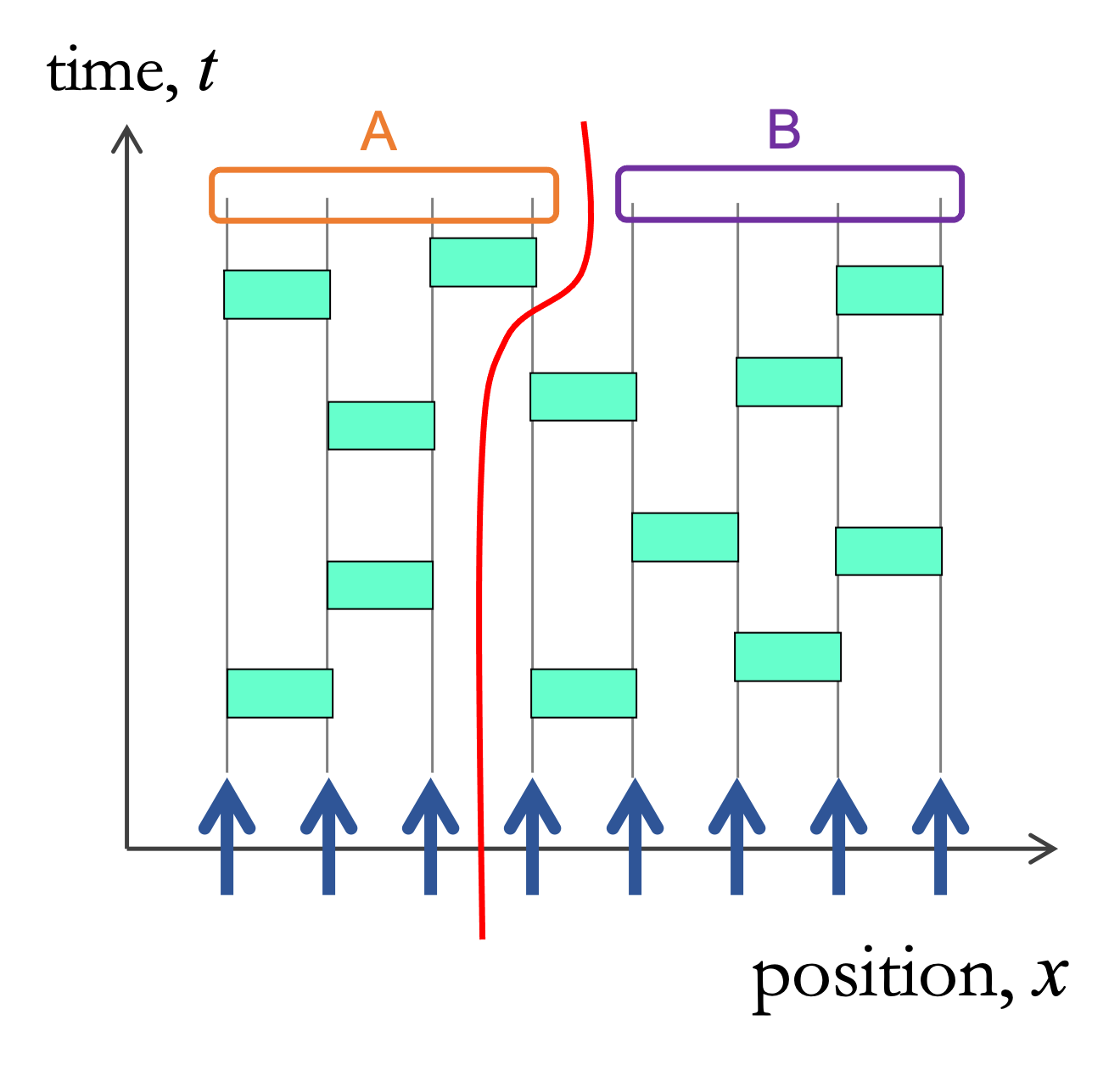}
\caption{The minimal cut partition that separates A and B.
} \label{fig:circuitmincut}
\end{figure}

In fact, this choice represents the \emph{minimal cut}. In this example there is no way to partition the network into two pieces that has a smaller number of cuts. So this minimal cut provides the tightest classical upper bound I can find on the entanglement:
\be 
S_n \leq N_\textrm{min cut} \times \log 2 = S_0.
\label{eq:Smincut}
\ee 
Actually, for the case of random unitary operators, it happens that $S_0$ is \emph{exactly} equal to $N_\textrm{min cut} \log 2$.

In this way the behavior of $S_0$ is transformed into a classical problem about the geometry of the quantum circuit. One only need know the physical structure of the circuit to learn something about the entanglement entropy and how it changes with time (and with the way I choose to partition my system into subsystems A and B).

\subsubsection{Universality class of entanglement growth and its fluctuations}

Now that we have mapped the behavior of $S_0$ on to a classical geometry problem, we can look around to see whether someone smarter than ourselves has already solved that problem. In this case it turns out that they have. At long times and large system sizes, the minimal cut problem for a circuit with randomly-placed unitaries can be mapped onto a classical problem in statistical mechanics known as the directed polymer in a random environment (DPRE). In the DPRE problem, one imagines laying a continuous polymer down on a random energy landscape in such a way that the polymer's energy is minimized. Here the ``polymer'' is the red line partitioning the circuit into two pieces and the ``energy'' is the value of $N_\textrm{cut}$, i.e.\ the number of circuit legs that are severed by the the minimal cut.

\begin{figure}[!h]
\centering
\includegraphics [width = 0.6\textwidth]{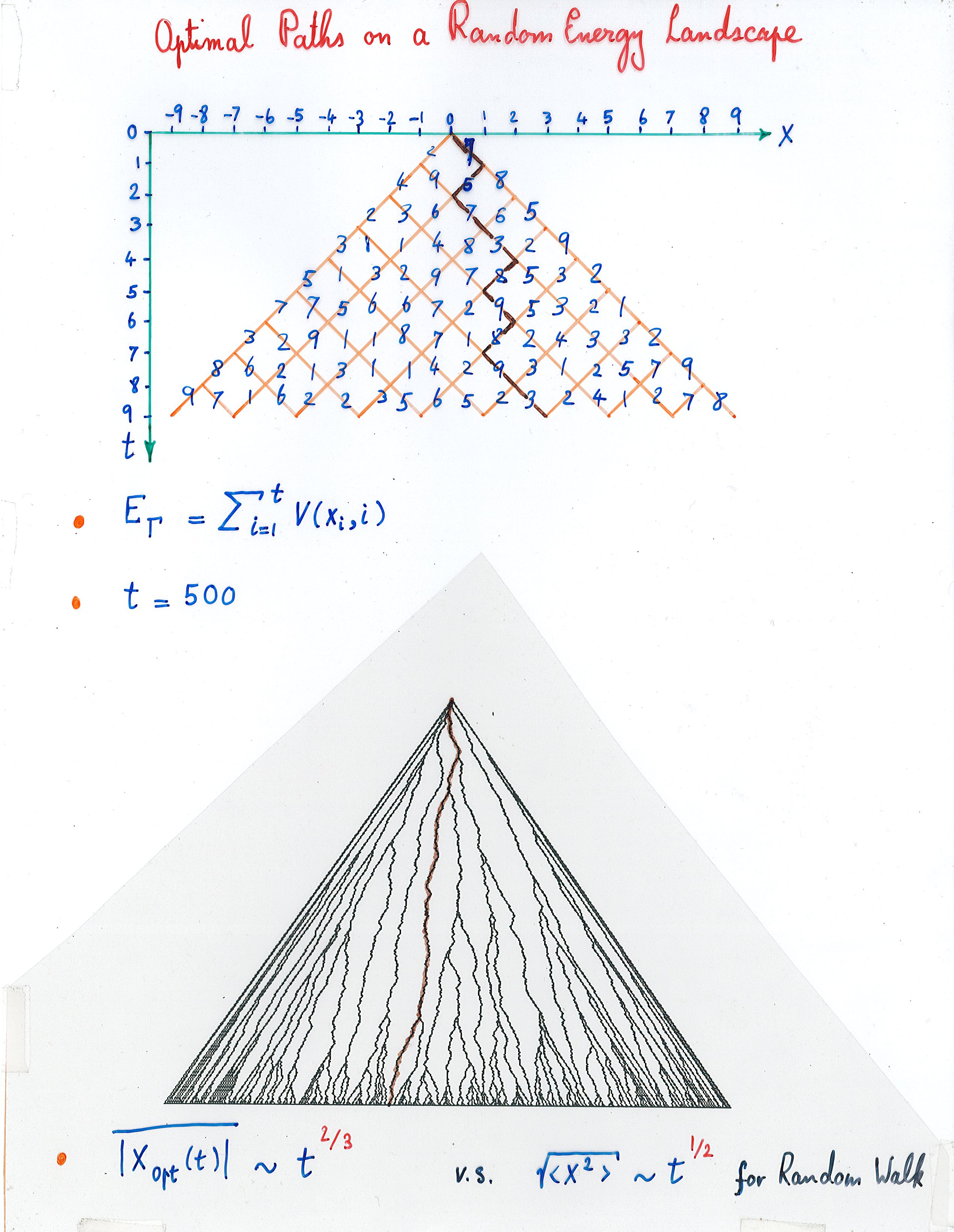}
\caption{An illustration of the DPRE problem, taken from Mehran Kardar's colloquium on the KPZ universality class (\href{https://www.mit.edu/~kardar/research/seminars/Growth/talks/Kyoto/introduction.html}{https://www.mit.edu/$\sim$kardar/research/seminars/Growth/talks/Kyoto/introduction.html}).
} \label{fig:DPRE}
\end{figure}

The beautiful thing about this analogy is that it allows one to figure out everything about the scaling and statistical fluctuations of the entanglement growth in a random circuit just by looking up the answers for the DPRE problem. As it happens, the DPRE problem belongs to the universality class of the Kardar-Parisi-Zhang (KPZ) equation.\footnote{The KPZ equation is commonly said to describe the evolution of the height $h$ of a randomly growing surface, like snow accumulating on a rooftop or the burning front of a piece of paper. It is generally written as $\partial_t h = \nu \partial_x^2 h - \beta (\partial_x h)^2 + \eta(x, t) + c$, where $\eta(x, t)$ describes random noise and the quantities $\nu$, $\beta$, and $c$ are parameters.}
From this mapping one can write down the scaling behaviors of the entanglement $S(x,t)$, where $x$ is the size of the left subregion (the distance from the left-hand side of the circuit to the point of separation between subsystems A and B) and $t$ is the evolution time. The primary results are:
\begin{enumerate}
    \item $S(x,t)$ grows linearly in time at short times, and then saturates to a constant value $S(x,t) \sim x$ for $t \gg x$. (At such late times the minimal cut exits through the lateral boundary of the circuit, while for $t \ll x$ it exits through the bottom boundary.)
    \item Statistical fluctuations in $S(x,t)$ grow with time as $\delta S \propto t^{1/3}$.
    \item Fluctuations in $S(x,t)$ have a correlation length $\xi$ in space that grows as $\xi \sim t^{2/3}$.
\end{enumerate}

\begin{figure}[!h]
\centering
\includegraphics [width = 0.6\textwidth]{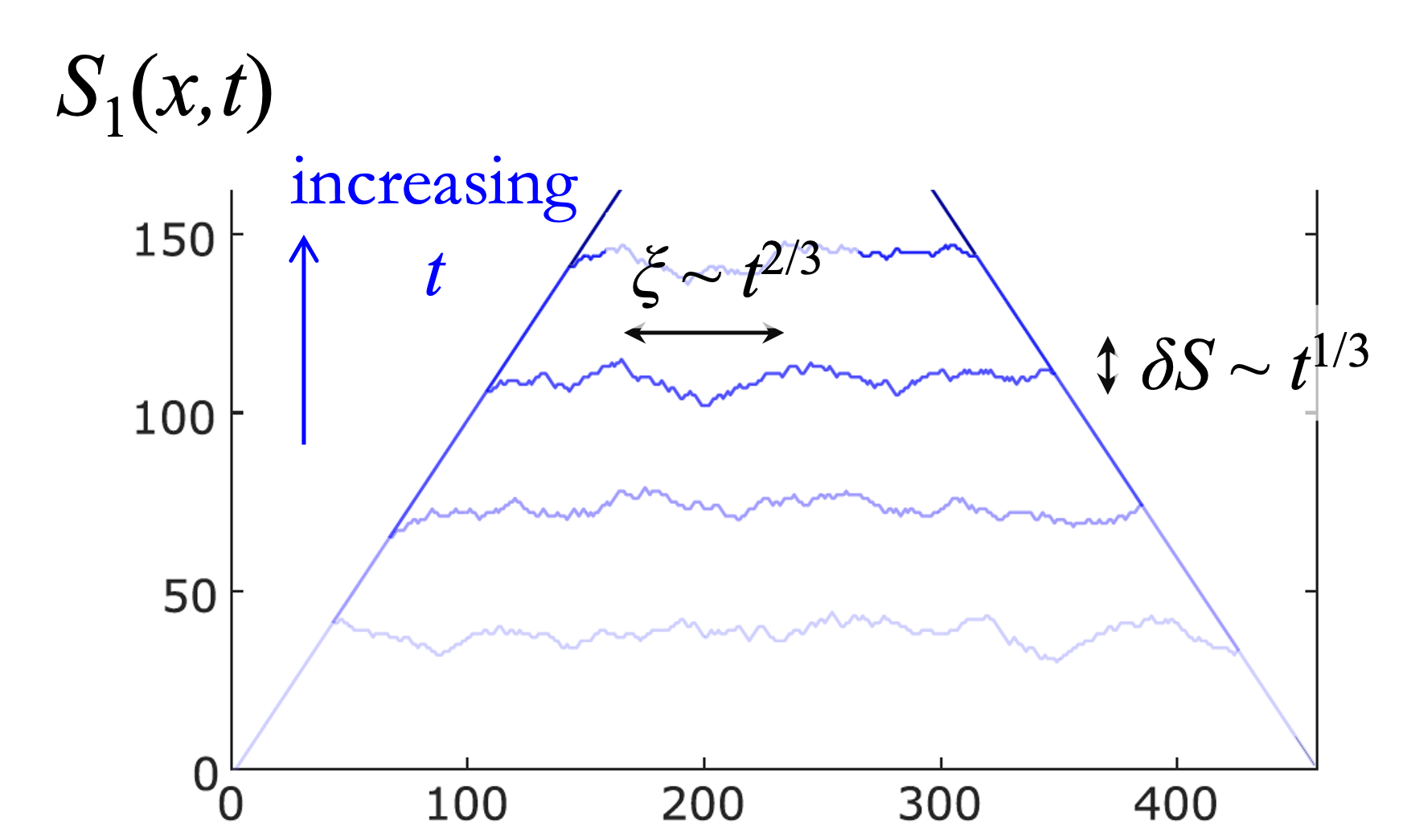}
\caption{Numerical simulations of the entanglement entropy $S_1(x,t)$ for a random Clifford circuit with 459 qubits. Adapted from Ref.~\cite{Nahum_Quantum_2017}.
} \label{fig:Sfluctuations}
\end{figure}

The realization that the dynamics of entanglement growth belongs to the KPZ universality class is surprisingly powerful. Once you've made this realization, you can look to any other system that belongs to the KPZ universality class and find a way to analogize it to the problem of quantum entanglement growth. In Ref.~\cite{Nahum_Quantum_2017}, the authors point out three different and unexpected analogies to entanglement growth (in addition to DPRE, they make analogies to classical surface growth and to the asymmetric simple exclusion process, which is often used to model traffic jams). Each analogy brings some unexpected insight to the problem of entanglement growth in a random circuit.

\subsection{The large-$q$ limit}
\label{sec:largeq}

I mentioned in \ref{sec:mincut} that the minimal cut generally provides a bound on $S_n$, and that it exactly gives the zeroth Renyi entropy $S_0$ (Eq.~\ref{eq:Smincut}). So the minimal cut generally gives an upper bound $S_0$ on the ``generic'' entanglement $S_n$ with $n \geq 1$. But there is no guarantee that $S_{n \geq 1}$ will be particularly close to $S_0$. It happens, though, that there is a particular limit where $S_{n}$ approaches $S_0$ exactly for all $n$. This is the limit where each qubit in the system is replaced by a $q$-state \emph{qudit} and I allow the number of states $q \rightarrow \infty$. In this limit, \emph{all} the Renyi entropies $S_n$ behave the same as the ``classical'' one $S_0$.

There is a proof that $S_n \rightarrow S_0$ at $q \rightarrow \infty$ in Ref.~\cite{Nahum_Quantum_2017}. But a heuristic way to rationalize this result is as follows. Consider two spin-$q$ qudits A and B that are initially unentangled and are then acted on by a single random unitary matrix. This unitary matrix takes the two spins to an entangled state $\ket{\psi}$ that can be described via a Schmidt decomposition (see Sec.\ \ref{sec:Schmidt})
\be 
\ket{\psi} = \sum_{i = 1}^q \sqrt{\lambda_i} \ket{i_A} \otimes \ket{i_B}.
\ee
The $n$th Renyi entropy is
\be 
S_n = \frac{1}{1 - n} \log \left( \sum_{i=1}^q \lambda_i^n \right).
\ee 
As long as the unitary matrix is random, then all $q$ of the Schmidt coefficients $\lambda_i$ will be nonzero with probability 1, which means that the zeroth Renyi entropy $S_0 = \log q$. In principle, to calculate the other Renyi entropies $S_n$ we need to know the values of the $\lambda_i$'s, which are random in some way, subject to the constraint that $\sum_i \lambda_i = 1$.  But the key is that, when $q$ is large, nearly all random states have sets of $\{\lambda_i\}$ such that most of the $\lambda_i$'s are of the same order of magnitude $\sim 1/q$. So in this large-$q$ limit we can estimate the sum
\be 
\sum_{i=1}^q \lambda_i^n \sim q \times \left(\frac{1}{q} \right)^n = q^{1-n}.
\ee 
From Eq.~\ref{eq:Sn} this gives exactly $S_n = \log(q) = S_0$.

So the point is that, when $q$ is large, the very large Hilbert space of each qudit guarantees that just about any randomly chosen state is very close to being maximally entangled. So as long as we are using random operators we can be confident that all $S_n$ will approach the upper bound $S_0$.

\section{The measurement-induced entanglement phase transition}
\label{sec:globalsymmetries}

The idea of the measurement-induced entanglement phase transition (which I will shorten to just ``measurement phase transition'', MPT) came from a very simple observation: unitary operators tend to increase the entanglement, while measurements tend to decrease it. For example, I can entangle two spins by applying a unitary operator to them, but if I measure the spin of either one I will disentangle them again. So what happens if I have a collection of many spins that is simultaneously being acted on by unitary operators and undergoing sporadic measurement? How is the competition between entangling forces (unitary operators) and disentangling forces (measurements) resolved? Do the unitary operators win so that the entanglement continues to grow with time? Or do the measurements win so that the entanglement growth is blocked?

The answer, perhaps surprisingly, is that there is a \emph{phase transition} between these two possibilities as a function of the measurement rate. When measurements are made sufficiently infrequently, there is an ``entangling phase'', in which entanglement grows and becomes extensive. When measurements are more frequent than a specific critical rate, however, the dynamics is ``disentangling'' and cannot support extensive entanglement.

There are different ways to characterize the entangling and disentangling phases (sometimes called the ``weak monitoring" and ``strong monitoring'' phases). A general summary of the features of the two phases is like this:
\begin{itemize}
    \item[] \textbf{Entangling phase:}
    \begin{itemize}
        \item \emph{Entanglement grows and becomes extensive.} \\ 
        Starting with a pure initial state having low entanglement, entanglement grows linearly in time and ultimately saturates at a value proportional to the subsystem size.
        \item \emph{An initially mixed state is basically never purified.} \\ 
        A highly mixed initial state remains highly mixed for a long time. Its mixedness decays to zero only over a time scale that is exponential in the system size.
    \end{itemize}
    \item[] \textbf{Disentangling phase:}
    \begin{itemize}
        \item \emph{Entanglement is blocked from growing.} \\ 
        An initially pure state with low entanglement will never develop extensive quantum entanglement. The growth rate of entanglement goes to zero in an order-1 time scale. A highly entangled initial state will see its entanglement drop to a value of order-1.

        \item \emph{Initially mixed states are easily purified.} \\
        The mixedness of a highly-mixed initial state decays to zero over an order-1 time scale, even for large system sizes.
        
    \end{itemize}
    
\end{itemize}

\begin{figure}[!htb]
\centering
\includegraphics [width = 0.65\textwidth]{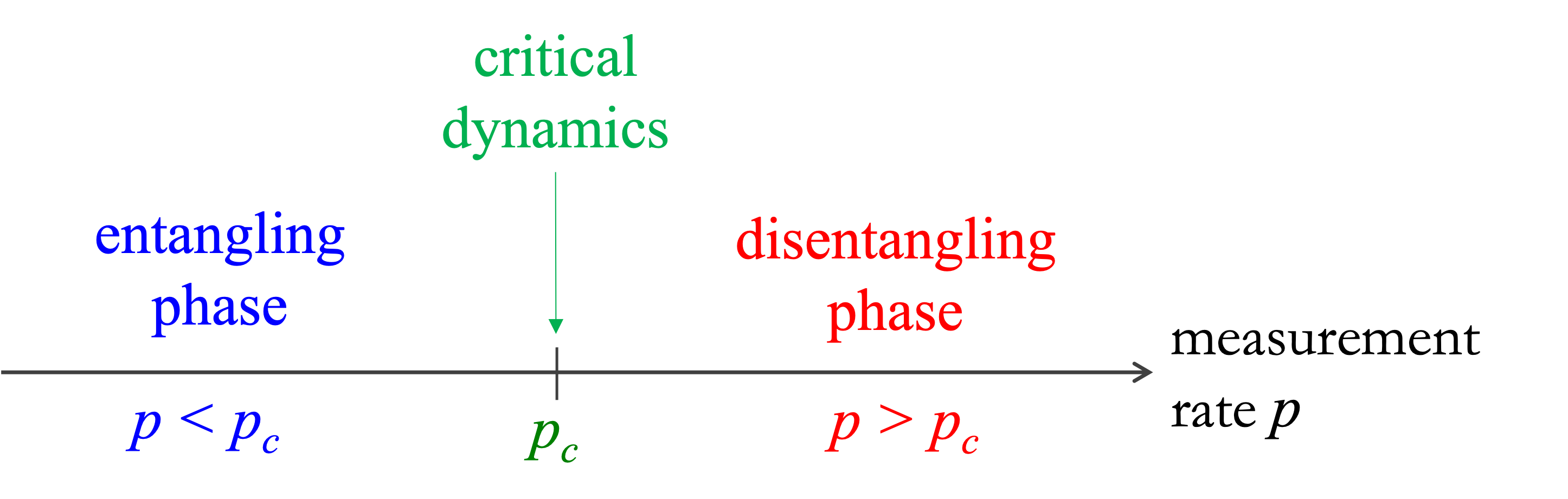}
\caption{The generic phase diagram of the measurement phase transition.
} \label{fig:phasediagram}
\end{figure}

How best to understand the measurement phase transition (MPT) is largely still an open question. Just 5 years after the MPT was first suggested, there are now many contexts in which an MPT or something like it is known to appear. But in most of these contexts there is still no exact solution for the transition or its critical properties.

But since I was personally involved in one of the first papers to suggest the existence of the MPT (Ref.~\cite{Skinner_Measurement_2019}, which was released simultaneously with Refs.~\cite{Li_Quantum_2018, Chan_Unitary-projective_2019} considering the same question), I can at least explain why we expected the MPT to exist.

\subsection{The original problem in $1 + 1$ dimensions}
\label{sec:1+1D}

The first context in which the MPT was identified is as follows. Consider a 1D line with $L$ spins/qubits that start out in a simple product state like $\ket{\uparrow \uparrow \uparrow ...}$ and evolve via a brickwork circuit of unitary operators. We can imagine that all the operators are chosen Haar-randomly.\footnote{This choice of Haar-random matrices turns out to not matter. Using Clifford unitaries or even taking every brick to be the same unitary matrix\footnotemark \ gives qualitatively the same results for the MPT.}\footnotetext{So long as the unitary chosen does not introduce any additional conserved quantities, like conserving the total $S_z$ of the two spins.\footnotemark}\footnotetext{I did not realize that you could nest a footnote inside a footnote.}
After each time step (defined as a single row of operators), there is a finite classical probability $p$ that any given spin will be measured.

\begin{figure}[!h]
\centering
\includegraphics [width = 0.8\textwidth]{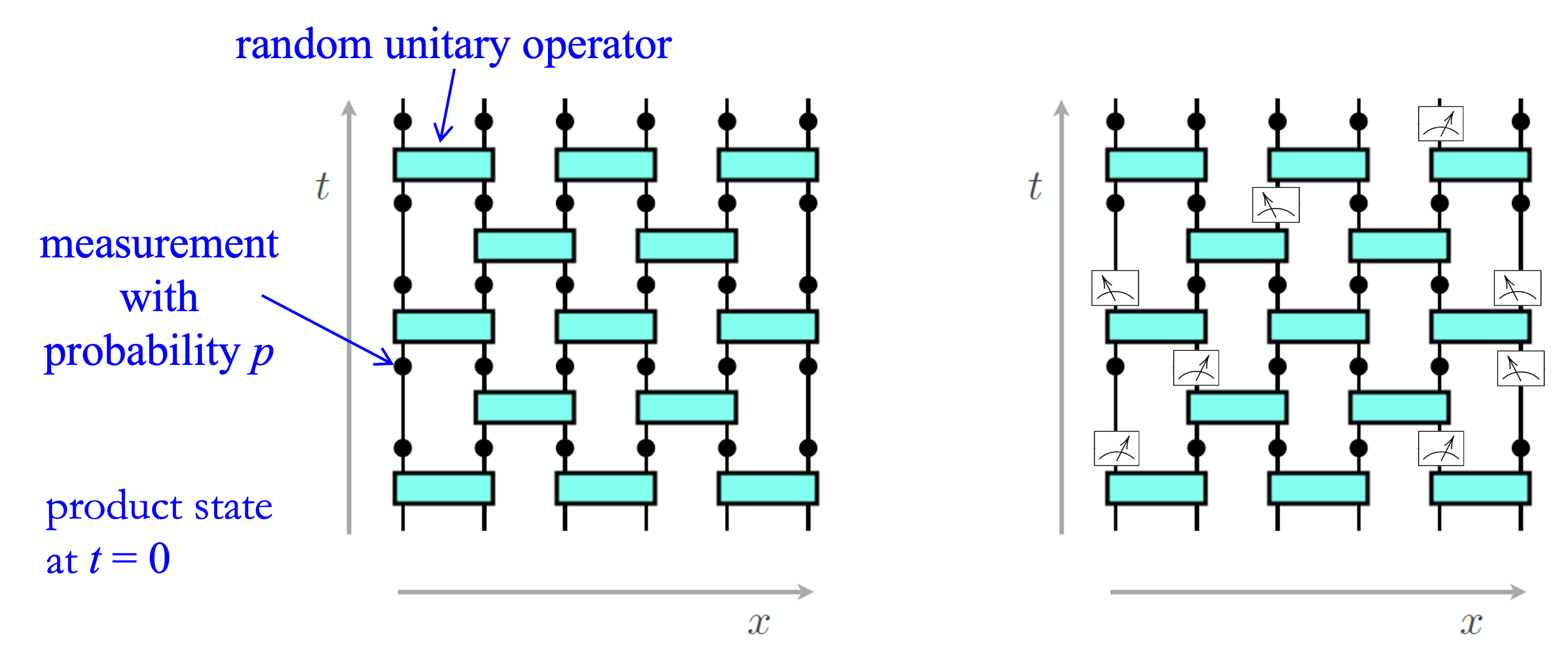}
\caption{A model for a quantum circuit with both unitary operators and measurements (sometimes called a ``hybrid circuit'' or ``monitored circuit''). The right side shows a specific realization, including a random choice for the times and locations of the measurements as well as a particular set of measurement outcomes.
} \label{fig:circuitwithmeasurements}
\end{figure}

In addition to randomly selecting the times and locations of the measurements, for a given realization of the circuit the measurement outcomes themselves are random. Each measurement yields either $\uparrow$ or $\downarrow$ for the corresponding spin, with probabilities that are given by the Born rule:
\be
P(\uparrow) = \left| \braket{\uparrow|\psi} \right|^2, \hspace{5mm}
P(\downarrow) = \left| \braket{\downarrow|\psi} \right|^2 = 1 - P(\uparrow),
\ee
where $\ket{\psi}$ indicates the wave function just before the measurement. To simulate a measurement, one of these two outcomes is chosen with the corresponding Born rule probability, and then the wave function is collapsed to the state consistent with the measurement outcome.\footnote{``Collapsing the state'' means applying the projection operator $\hat{P}_{\uparrow/\downarrow} = (1 \pm \hat{\sigma_z})/2$ to $\ket{\psi}$ and then renormalizing it.}

After this protocol has been followed for a certain time $t$, one can ask how entangled the state $\ket{\psi(t)}$ is. Of course, defining entanglement always requires one to partition the system into two subsystems A and B -- the most common choice is to partition the system into its left and right halves, in which case refer to the entanglement entropy as the ``bipartite entanglement''.

In terms of this bipartite entanglement entropy $S$ (I am being intentionally vague, for now, about the order $n$ for the Renyi entropy), the phenomenology of the transition is as follows. In the limit $L \rightarrow \infty$ but finite time $t$,
\be 
S(t, L \rightarrow \infty) \propto
\begin{cases} 
      t, & p < p_c \\
      \log t , & p = p_c \\
      \textrm{const.}, & p > p_c 
   \end{cases}.
\ee
So roughly speaking, the critical measurement rate separates a phase where entanglement grows from a phase where entanglement doesn't grow. Exactly at the critical point, the entanglement technically grows, but in a stupid way (logarithmically).

In the limit $t \rightarrow \infty$ but finite system size $L$, the behavior of the bipartite entanglement entropy is
\be 
S( L, t \rightarrow \infty) \propto
\begin{cases} 
      L, & p < p_c \\
      \log L , & p = p_c \\
      \textrm{const.}, & p > p_c 
   \end{cases}.
\ee
So we often say that the critical measurement rate separates a ``volume law'' phase (where the entanglement entropy is proportional to the system size) from an ``area law'' or ``boundary law'' phase (where the entanglement entropy is proportional to the boundary between the two subregions, which for a 1D system is just 1 site).

\subsection{The postselection problem}
\label{sec:postselection}

It is tempting to think that the MPT problem describes a system evolving under quantum dynamics while also interacting with an environment (the measurer). In this sense it sounds excitingly generic, since all real quantum systems involve interactions (perhaps unwanted) with an environment. We normally call such interactions ``decoherence'': they tend to take the quantum state away from something we know toward something we don't know (more on this in a moment). So it is tempting to think that the existence of an MPT implies something ubiquitous about what happens to a quantum system coupled to an external environment.

But there is a very big difference between \emph{measurements} of the kind that are studied in the MPT problem and \emph{decoherence} of that kind that always plagues quantum systems. Namely, a measurement involves projecting the wave function into a known state, consistent with the outcome of the measurement. So after a measurement, a pure state (described by a specific wave function, $\rho = \ket{\psi} \bra{\psi}$) remains a pure state. But during decoherence, a pure state is taken to a mixed state, a statistical sum of different wave functions that correspond to different possible measurement outcomes. In effect, a decoherence event is like a measurement whose outcome is unknown and must be averaged over. So the resulting density matrix after an instance of decoherence is like an average over two different density matrices, each associated with one of the two possible measurement outcomes.

For this reason the MPT is a feature only of the \emph{post-selected state}, and not of the \emph{measurement-averaged state}. What I mean is that, for a given set of measurement outcomes $\{m_i\}$, I can construct the density matrix
\be 
\rho_{\{m_i\}} = \ket{\psi_{\{m_i\}}} \bra{\psi_{\{m_i\}}}.
\ee 
From this density matrix $\rho_{\{m_i\}}$ I can define the entanglement entropy $S$ for a given realization of the quantum circuit. With $S$ defined in this way, I will be able to observe the MPT as a function of the measurement rate $p$ (and I can safely average $S$ over different realizations of the circuit for a given $p$).

But the MPT is \emph{not} evident in the measurement-averaged density matrix
\be 
\rho_\textrm{avg} = \sum_{\{m_i\}} p(\{m_i\}) \ket{\psi_{\{m_i\}}} \bra{\psi_{\{m_i\}}}
\ee 
(where $p(\{m_i\})$  represents the probability of a particular set of measurement outcomes). In general $\rho_\textrm{avg}$ flows toward some kind of boring infinite-temperature equilibrium state after many measurements (i.e., $\rho_\textrm{avg}$ becomes something like a diagonal matrix with uniform entries $1/2^L$ along the diagonal). It is completely featureless at the critical point $p_c$.

This inadequacy of $\rho_\textrm{avg}$ for seeing the MPT is sometimes called the \emph{postselection problem}. What it implies is that one cannot simply couple a quantum system to a bath and expect some interesting critical properties to appear as a function of increasing coupling strength. The MPT is a property of the \emph{postselected quantum state} -- it is a feature of the entanglement entropy, which is derived from a specific pure quantum state, and cannot be deduced once we have averaged the state over different measurement outcomes.

Because of the postselection problem, observing the MPT experimentally is difficult. In order to deduce the entanglement entropy of a particular quantum state, one must know the state (i.e., the density matrix). But in order to deduce the density matrix from measurements, one needs to prepare the same quantum state many times. And this means not just running the same circuit many times, but achieving the exact same record of measurement outcomes many times. Since the measurement outcomes are intrinsically random (they are given by the Born rule), preparing the same state multiple times is a problem that is exponentially difficult in the number of measurements. In a random circuit, the probability of getting a specific measurement record is $2^{-(\textrm{\# of measurements})}$. 

There have nonetheless been some heroic efforts to observe the MPT experimentally in quantum simulators (e.g., \cite{Noel_Measurement_2022, Koh_Experimental_2022}). But these experiments have generally been limited to small systems and short times precisely because of the severity of the postselection problem. Finding ways to get around the postselection problem is a serious ongoing research direction.

\subsection{The minimal cut and mapping to percolation}

So how does one understand the origin of the MPT? There are different ways to rationalize its existence, but in the case of the 1+1D circuit we anticipated the existence of the MPT using arguments based on the minimal cut.

\begin{figure}[!h]
\centering
\includegraphics [width = 0.8\textwidth]{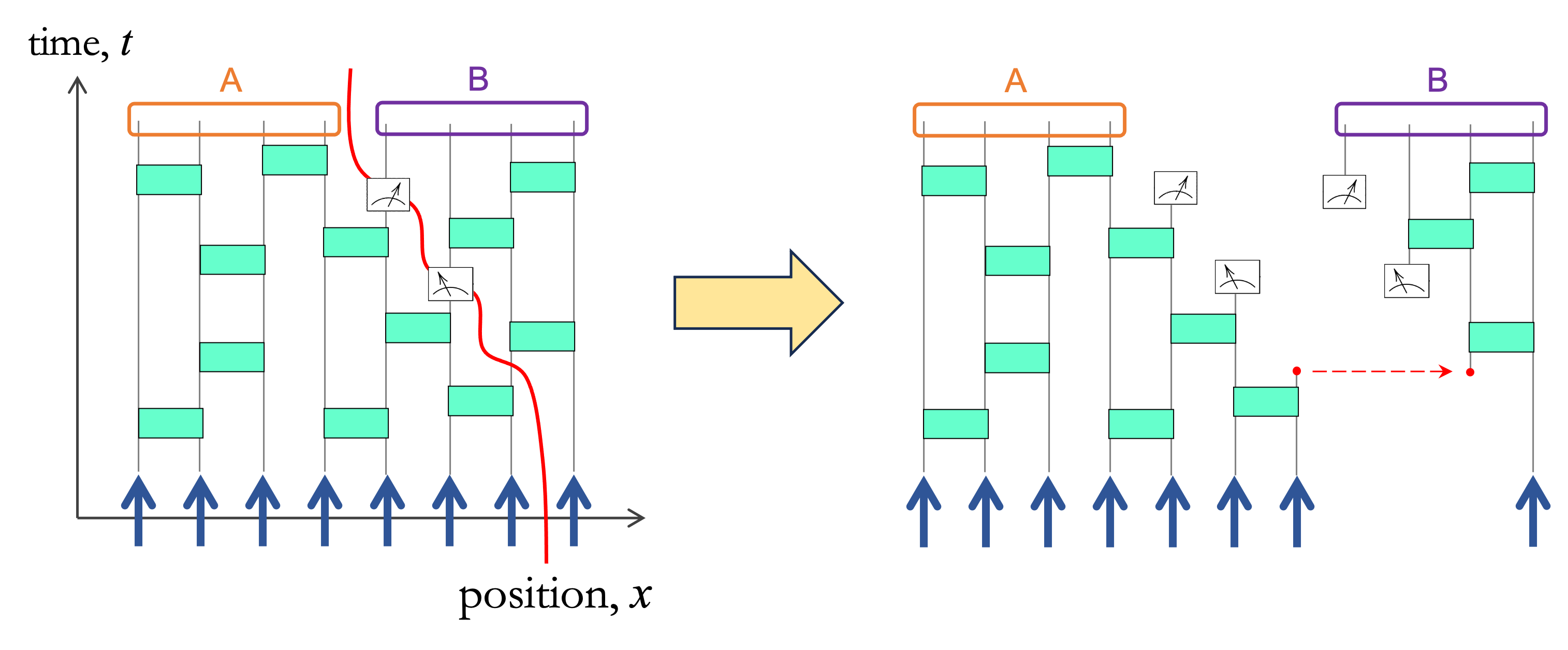}
\caption{This example is identical to Fig.~\ref{fig:circuitwcut}, except that two measurements have been added to the circuit, which the red-line cut passes through.
} \label{fig:circuitcut-withmeasurement}
\end{figure}

The key idea is to realize that, for the purposes of the minimal cut, a measurement acts like a kind of ``pre broken link" in the circuit. Consider, for example, a circuit that is identical to the one in Fig.~\ref{fig:circuitwcut} except that it contains two mid-circuit measurements, as shown in Fig.~\ref{fig:circuitcut-withmeasurement}. These measurements constitute points in time and space where I know precisely the value of the spin: at such moments the state of the spin is disconnected from its past, in the sense that I don't need to know the previous state in order to know the value of the spin at that moment. So if I draw my cut in such a way that it passes through the location of those two measurements, then I can simply feed the known state of the spin into the right-side tensor network. There is no longer any information transmitted by those two points between the input state of the left tensor network and the output state of the right tensor network. In the language of the decomposition in Eq.~\ref{eq:cutdecompositionrearranged}, the existence of measurements along the cut means that the decomposition has fewer terms, because the measured spins have definite values and their state is not summed over. Consequently the amount of information transmitted between these two tensor networks is at most \emph{one bit}, and not 3 as it was before.

So now the nature of the ``minimal cut game" is modified slightly. In the absence of measurements, the game was: 
\begin{displayquote}
    Find a line that starts at the midpoint between A and B and exits the circuit while passing across the smallest possible number of circuit legs.
\end{displayquote} 
But now the game is:
\begin{displayquote}
    Find a line that starts at the midpoint between A and B and exits the circuit while passing across the smallest possible number of circuit legs. \emph{Locations of measurements act like pre-broken legs and can be traversed for free.}
\end{displayquote}

This revised minimal cut game is in fact an instance of a different well-studied classical stat mech problem called \emph{first passage percolation}. In first passage percolation, one attempts to draw a line from the top of a network to the bottom, keeping in mind that some passages in the network are blocked.\footnote{To my knowledge, this picture is why we refer to this class of geometric problems as ``percolation": people imagined it as a problem of water filtering through a disordered porous medium like coffee grounds.}

\begin{figure}[!h]
\centering
\includegraphics [width = 0.9\textwidth]{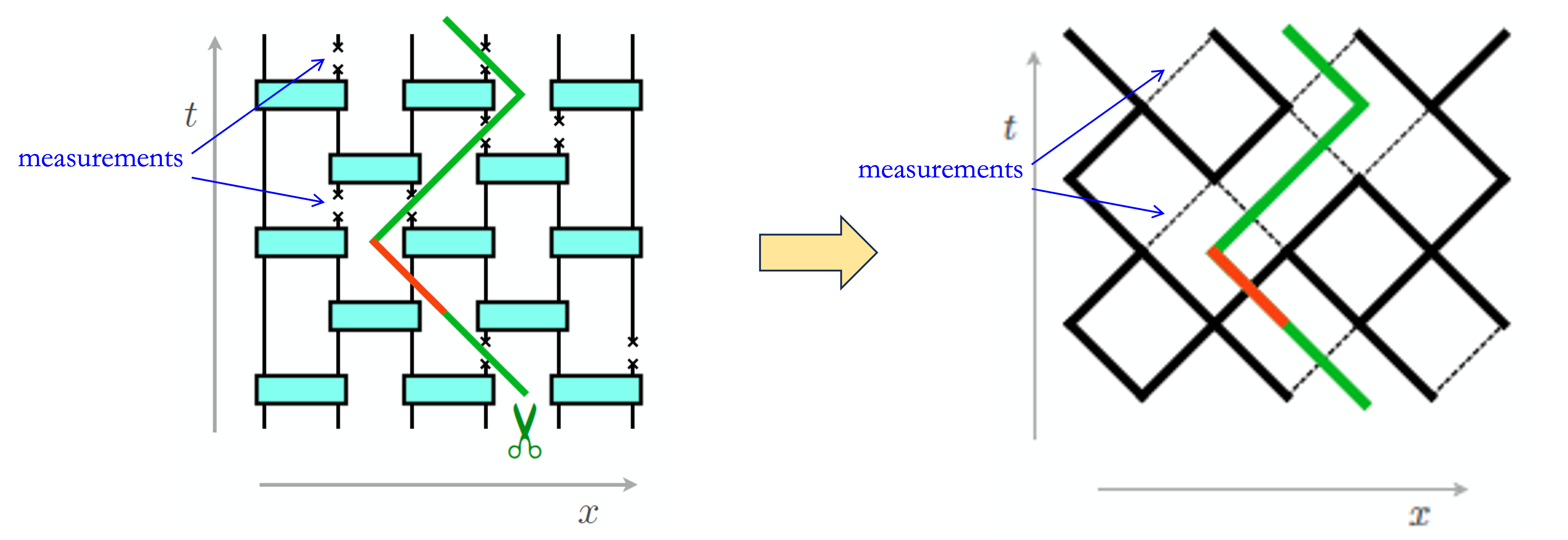}
\caption{The mapping between the ``minimal cut'' problem in a circuit with measurements and the problem of first-passage percolation in a random environment.
} \label{fig:percmapping}
\end{figure}

To understand the resulting behavior of the minimal cut, one need only know these facts about percolation in random networks:
\begin{itemize}
    \item At $p < p_c$, the network resembles a ``ripped up fisherman's net'', with big holes of size $\xi \sim 1/(p_c - p)^\nu$, which grow larger as one approaches the transition ($\nu > 0$ is the correlation length critical exponent). Those big holes in general have many dangling loose ends, but these are irrelevant for the minimal cut. The minimal cut therefore requires making roughly one cut per length scale $\xi$.
    \item Exactly at $p = p_c$, the correlation length $\xi$ diverges and the network has no characteristic length scale. One can say that the network resembles a scale-free distribution of ``empty chambers'' of all length scales. The minimal cut involves passing from a small-sized chamber (which is generically where the cut begins) to ever larger length scales until one finds an empty chamber comparable in diameter to the system size.
    \item At $p > p_c$, the network falls apart into many disconnected pieces with characteristic size $\xi \sim 1/(p - p_c)^\nu$. The minimal cut may begin inside such a piece, but after exiting that piece the cut can proceed freely to the boundary of the network.
\end{itemize}

\begin{figure}[!h]
\centering
\includegraphics [width = \textwidth]{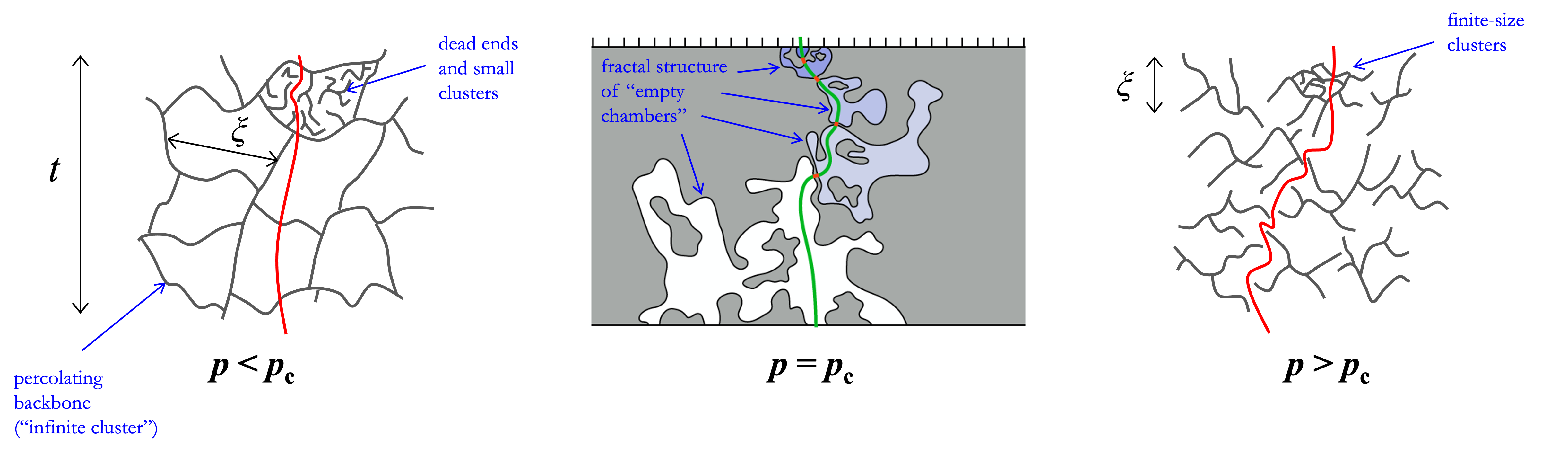}
\caption{Pictures of the minimal cut in a random network for different measurement regimes.
} \label{fig:network_percolation}
\end{figure}

Think about these facts a bit (and perhaps stare at Fig.~\ref{fig:network_percolation}), and you'll reach the following conclusions about how the entanglement behaves (say, as a function of time in the limit of infinite system size):
\be 
S(t, L \rightarrow \infty) \sim
\begin{cases} 
      t \times (p - p_c)^\nu, & p < p_c \\
      \log t , & p = p_c \\
      -\nu \log(p - p_c), & p > p_c 
   \end{cases}.
\label{eq:Stcases}
\ee
In these cases the minimal cut exits through the bottom of the circuit (the sides are far away, given $L \rightarrow \infty$). Importantly, the entanglement growth rate $dS/dt$ vanishes continuously as one approaches the transition $p = p_c$. In this sense the MPT is like a continuous, second-order transition rather than a first-order one.

In the limit of long time and finite system size, the minimal cut exits through one of the side boundaries of the circuit (the bottom time boundary is far away since $t \rightarrow \infty$). But the scaling with system size $L$ is similar to Eq.~\ref{eq:Stcases}:
\be 
S(L, t \rightarrow \infty) \sim
\begin{cases} 
      L \times (p - p_c)^\nu, & p < p_c \\
      \log L , & p = p_c \\
      -\nu \log(p - p_c), & p > p_c 
   \end{cases}.
\label{eq:SLcases}
\ee
Thus at long times it is the \emph{entanglement density} $S/L$ that vanishes as one approaches $p = p_c$ from below. In other words, the entanglement goes from volume-law to area-law.

Notice (if you want) that all three of these regimes can be captured by the following critical scaling forms:
\be 
S(t, p, L \rightarrow \infty)  - S(t, p=p_c, L \rightarrow \infty) = F \left( (p-p_c) t^{1/\nu} \right)
\label{eq:St}
\ee 
and 
\be 
S(L, p, t \rightarrow \infty)  - S(L, p=p_c, t \rightarrow \infty) = G \left( (p-p_c) L^{1/\nu} \right),
\label{eq:SL}
\ee
where $F(x)$ and $G(x)$ are some unknown functions. Demonstrating this form of critical scaling for numerical simulation data was an important piece of evidence for the existence of the MPT.

Now, remember (from Sec.~\ref{sec:Sdef}) that the minimal cut only gives $S_0$, which is an upper bound on the ``generic'' Renyi entanglement entropies $S_{n \geq 1}$. The minimal cut argument shows that $S_0$ must have a transition, and in fact the properties of this transition for $S_0$ are known exactly from classical percolation: $p_c = 1/2$ and $\nu = 4/3$.
But since $S_0$ is only an upper bound for the generic case, the existence of a transition for $S_0$ does not immediately imply that $S_{n \geq 1}$ must have a transition (except in the large local Hilbert space limit $q \rightarrow \infty$, discussed in Sec.~\ref{sec:largeq}). It would be logically consistent if all $S_{n \geq 1}$ belonged to the disentangling phase. 

\begin{figure}[!h]
\centering
\includegraphics [width = 0.7\textwidth]{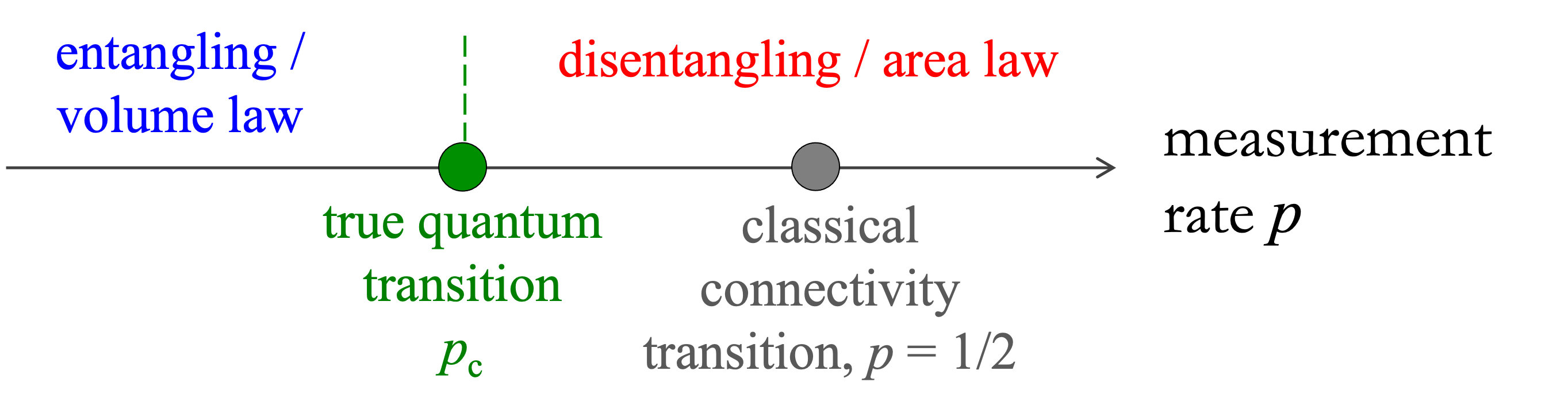}
\caption{Distinguishing two distinct values of the measurement rate: the classical connectivity transition $p = 1/2$, where the circuit falls apart into disconnected  pieces, and a smaller value $p_c$ for which the generic entanglement entropies $S_{n \geq 1}$ exhibit the MPT.
} \label{fig:pq_and_pc}
\end{figure}

But the intuition from the minimal cut (Eqs.~\ref{eq:Stcases} and \ref{eq:SLcases}) made it clear what phenomenological behavior to look for in numerical simulation studies. And at this point there is ample numerical evidence that Eqs.~\ref{eq:St} and \ref{eq:SL} are obeyed even for the generic case $n \geq 1$, for some values of $p_c$ and $\nu$ that are not, in general, equal to the ones given by classical percolation. So we have the scenario summarized in Fig.~\ref{fig:pq_and_pc}. Frustratingly, in the 1+1D circuit problem we have an exact solution for the ``classical connectivity" transition, but there is no exact solution as of yet for the location or critical properties of the generic transition.

\subsection{Mutual information and its classical interpretation}

In addition to looking at the bipartite entanglement entropy, a useful quantity for studying the MPT is the \emph{mutual information} between two sets of spins $a$ and $b$ that are well separated from each other (i.e., $a$ and $b$ are not next to each other in space). The mutual information is defined as\footnote{The $a$ and $b$ in Eq.~\ref{eq:mutualinfo} are not the same as the $a$ and $b$ in Eq.~\ref{eq:cutdecomposition} and you're going to have to deal with that.}
\be 
I_n(a,b) = S_n(a) + S_n(b) - S_n(a \cup b).
\label{eq:mutualinfo}
\ee 
Here, $S_n(a)$ represents the entanglement entropy between $a$ and everything outside of $a$. In the language of the minimal cut, $S_n(a)$ represents the cost of making a cut that removes $a$ from the rest of the circuit, $S_n(b)$ represents the same quantity for $b$, and $S_n(a \cup b)$ represents the cost of removing both sets $a$ and $b$ together.

\begin{figure}[!h]
\centering
\includegraphics [width = 0.5\textwidth]{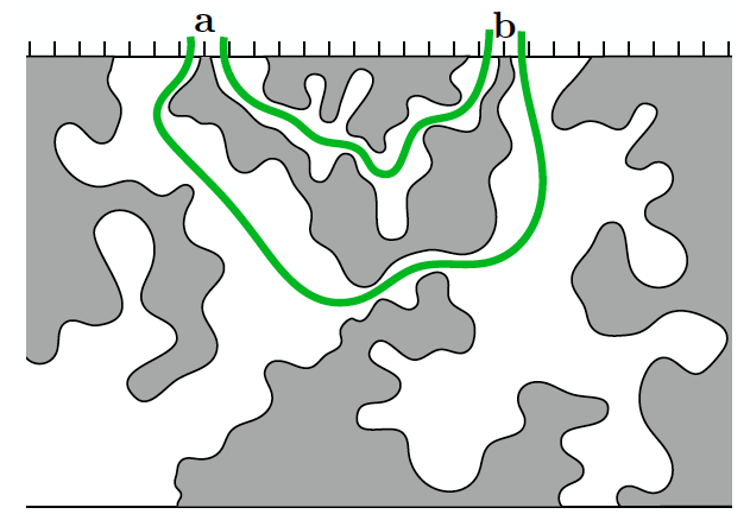}
\caption{An example configuration of the circuit for which removing $a \cup b$ is easier than removing $a$ and $b$ separately. The white space represents regions of the circuit with measurements and the gray space represents regions without.  
} \label{fig:mutualinfo}
\end{figure}

In most situations where $a$ and $b$ are far apart, the simplest way to remove $a \cup b$ from the rest of the system is to make a cut around $a$ and a cut around $b$. But in such situations we have $S_n(a \cup b) = S_n(a) + S_n(b)$, and the mutual information is zero. Thus,  in order for the mutual information $I_n(a,b)$ to be nonzero, it must be easier (in the minimal cut sense) to remove $a$ and $b$ together than to remove each one separately. At either $p < p_c$ or $p > p_c$, such a possibility is reasonably likely only if $a$ and $b$ are within one correlation length $\xi$ of each other, so that there is a chance of them belonging to the same ``connected cluster'' of the network (as depicted in Fig.~\ref{fig:mutualinfo}).

Thus, if the sets $a$ and $b$ are well separated, then the mutual information $I_n(a,b)$ has a \emph{peak} as a function of $p$ when $p \rightarrow p_c$, since that is where $\xi$ becomes long. Looking for a peak in the mutual information between two distant sets of spins is therefore a good method for identifying the location $p_c$ of the MPT.

\subsection{The entanglement transition as a purification transition}

So far I have been talking about the MPT as something that is reflected in the entanglement entropy of a pure state. But it turns out that there is a different, equivalent way of defining the transition: by asking whether a highly mixed initial state is purified by the dynamics.

A ``mixed state'' means a density matrix that cannot be written as the outer product of any wave function with itself, or in other words that cannot correspond to any specific wave function. Instead, a mixed state corresponds to a statistical sum of many different wave functions, like a random ensemble of different quantum states. The ``maximally mixed state'' refers to a density matrix that is proportional to the identity matrix, e.g.,
\be 
\rho_\textrm{maximally mixed} = \frac{1}{2^L} 
\begin{pmatrix}
1 &  &  & &  \\
 & 1 &  &  &\\
 &  & 1 &  & \\
 &  &  & \ddots & \\
 & & & & 1
\end{pmatrix}.
\ee 
This maximally mixed state has a large amount of classical entropy (``von Neumann entropy''):
\be 
S_{vN} = - \textrm{Tr} \left( \rho \log \rho \right).
\ee 
(This quantity $S_{vN}$ should not be confused with the \emph{entanglement entropy}, which is defined via a reduced density matrix, i.e., one for which part of the system has been traced out.)
For the maximally mixed state, $S_{vN} = L \log 2$. 
In a sense, saying ``I have a maximally mixed initial state'' is equivalent to saying ``I have absolutely no idea what the initial state is."

By contrast, any pure state has $S_{vN} = 0$. Pure states have no classical entropy because they correspond to some specific wave function, or in other words any for any pure state there exists some basis for which the density matrix can be written
\be 
\rho_\textrm{pure} = 
\begin{pmatrix}
1 &  &  & &  \\
 & 0 &  &  &\\
 &  & 0 &  & \\
 &  &  & \ddots & \\
 & & & & 0
\end{pmatrix}.
\ee 

Consider a circuit with unitaries and measurements that acts on a maximally mixed initial state. One can generally expect two possibilities: either the measurements are strong enough to purify the state, so that $S_{vN}/L$ goes to zero with time, or they are not, so that $S_{vN}/L$ remains finite. Gullans and Huse \cite{Gullans_Dynamical_2020} showed that there is in fact a phase transition between these two possibilities, and that it is the same MPT that controls the behavior of the entanglement entropy.

Proving the equivalence between the MPT and the purification transition is not especially simple, but again one can rationalize its existence using minimal cut thinking. Think of a circuit as a (nonunitary) time evolution operator that transforms an input state into an output state. A measurement acts like a ``hole'' in this time evolution operator: a moment where the state of the spin is known definitely without reference to its prior history. If these holes percolate horizontally across the circuit, or in other words if it is possible to draw a red line that separates the input and output of the circuit from each other without passing through any unbroken legs of the circuit, then the output of the circuit can be deduced without knowing the input. In other words, if there are enough measurements, then the input state is completely forgotten: the output can be fully reconstructed just by knowing the measurement outcomes and the unitary operators. 

\begin{figure}[!h]
\centering
\includegraphics [width = 0.35\textwidth]{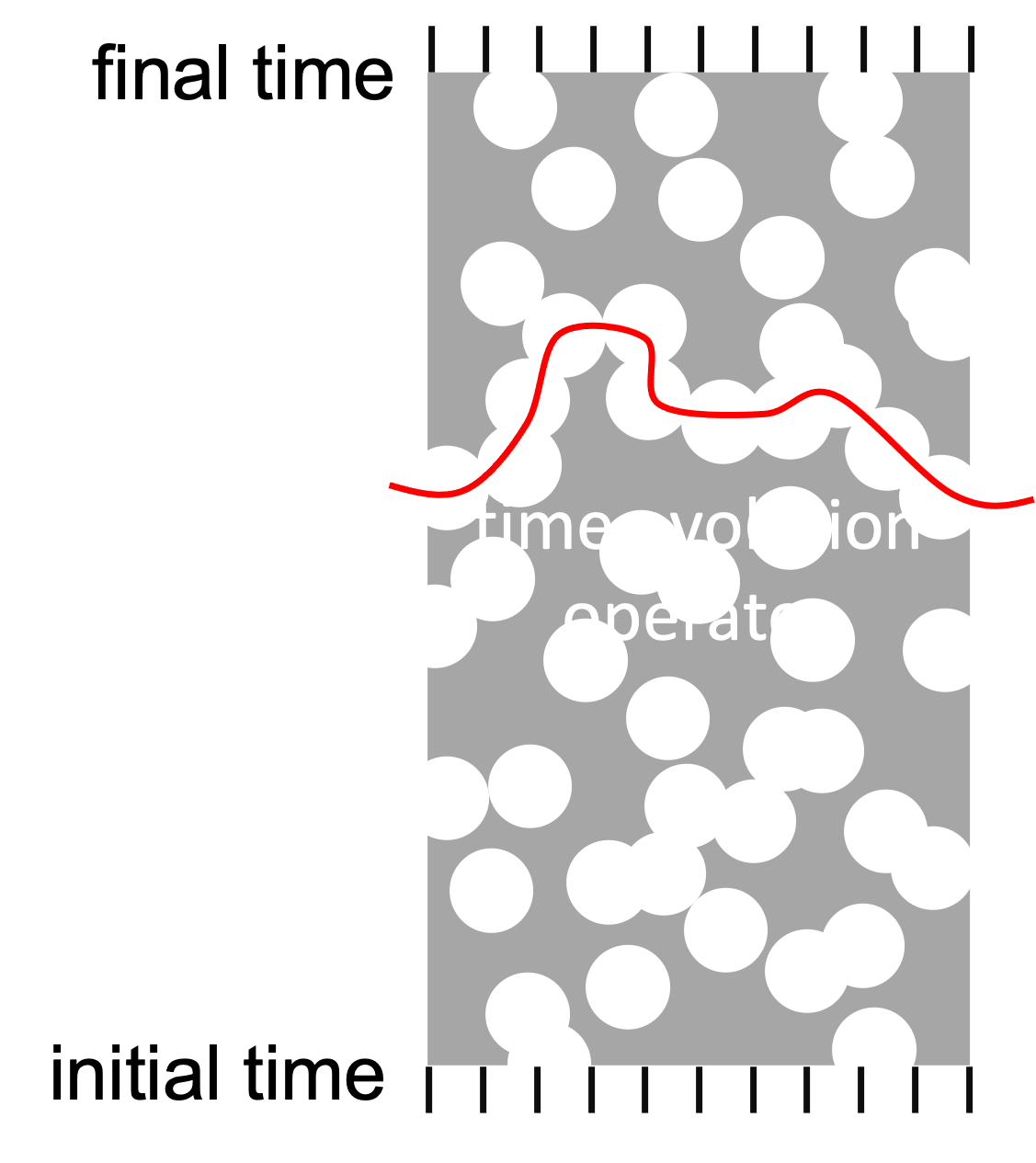}
\caption{ Measurements act like holes in the time evolution operator (the circuit), and if there are enough of them then one can disconnect the initial and final time states completely.
} \label{fig:circuit_with_holes}
\end{figure}

If the initial time state is forgotten, then the state reconstructed from the measurements is necessarily a pure state. So a maximally mixed initial state is purified. In the minimal cut picture, whether such a horizontal cut exists is controlled by the same percolation transition that determines the behavior of the entanglement entropy of a pure state, so it corresponds to the same MPT.

Of course, for any finite system, there is always a probability that the system will be purified by some unlikely choice of measurement locations, even if $p$ is small. For example, there is always a nonzero probability that I will happen to choose to measure every spin at the same time step. Such a choice is exponentially unlikely, but if it happens then it will instantly purify the state. So the general result is that the entropy (mixedness) of the state always decays to zero if I'm willing to wait long enough,
\be 
S_{vN}(t) \sim L e^{-t / \tau},
\ee 
but the decay time $\tau$ behaves very differently in the two phases:
\begin{alignat}{3}
      \log \tau & \propto && \ N, && \quad p < p_c \\
      \tau & \sim && \ O(1), && \quad p > p_c .
\end{alignat}
In this sense a large system is ``quickly purified'' at $p > p_c$, and ``basically never purified'' at $p > p_c$.

\subsection{What is it good for?}

The MPT is a new kind of phase transition: a transition in the dynamics of quantum entanglement, rather than in any thermodynamic quantity or the expectation value of any operator. But is it good for anything? Does it have any important implications for other physics that we care about?

The answer to this question so far is mostly: we don't know. One unambiguous implication of the MPT concerns the question of how difficult it is to simulate, on a classical computer, a quantum system coupled to an environment. As explained in Sec.~\ref{sec:postselection}, quantum systems decohering to an environment are ubiquitous, but they do not exhibit the MPT since decoherence is like a measurement whose outcome is unknown and therefore does not preserve the purity of the state. But a convenient way of \emph{simulating} decoherence is to randomly choose\footnote{according to the Born rule, of course} at each instant of decoherence a specific measurement outcome, and then follow the resulting pure state. One can then estimate the time-dependence behavior of some observable by following many such ``quantum trajectories'' and averaging the corresponding expectation value of the observable. This method is often called the method of ``quantum trajectories'' or ``quantum jumps" \cite{daley2014quantum}.

But for this computational method it is critical to know whether the pure state wave function that is being followed in its trajectory is highly entangled. If the state is highly entangled, then it is computationally very difficult to deal with. But if the state does not have extensive entanglement then it can be dealt with efficiently using some compact representation like a matrix product state. The MPT implies that there is a sharp phase transition in the amount of entanglement as a function of the rate of measurement in the trajectory. So the upshot is that the computational difficulty of describing a large quantum state with decoherence has a phase transition as a function of the rate of decoherence.

Other possible uses for the MPT are less direct, or more aspirational. For example, there is a sense in which the MPT can be viewed as a transition in the effectiveness of a randomly chosen quantum error-correcting code (see, e.g., \cite{Choi_Quantum_2020, Fan_Self-organized_2021, Gullans_Dynamical_2020}). The MPT also might have implications for the technique of ``classical shadow tomography'', in which properties of a quantum state are inferred by randomly making measurements on the state. There may be a sense in which making measurements exactly near the critical point of the MPT maximizes the effectiveness of the technique \cite{Hu_classical_shadow_2023, garratt2023probing}. Whether these connections give something practically useful for quantum computing remains to be seen.

%\subsection{Other contexts for the entanglement transition}

%I have not made any effort here to cite people thoroughly or fairly. So it's the least I can do to at least mention some other contexts where the MPT is being studied.

%First, the 1+1D problem can be generalized straightforwardly to higher spatial dimensions, for example a quantum circuit acting on a 2D or 3D lattice of spins. In this case one can extend the minimal cut idea to talk about a ``minimal membrane'' that partitions this higher-dimensional circuit into separate pieces. Since percolation exists as a geometric transition in any spatial dimensions (larger than 1), one can generically expect that the MPT exists in any number of spatial dimensions (including the infinite dimensional limit, in which all spins in a finite-sized system are nearest neighbors of each other).

%One can also abstract away from circuits with spatial locality to ask about a broader class of tensor networks, including those with a tree structure (which relate the state of many spins at the base of the tree to a single spin at the apex). Some of these tree tensor networks admit an exact solution for the generic MPT.

\newpage

{\small
\bibliography{reference.bib}
\bibliographystyle{utphys}
}

\end{document}